\documentclass[longauth]{aa}  
\usepackage{graphicx}
\usepackage{xcolor}
\usepackage{txfonts}
\usepackage{hyperref}
\usepackage{xspace}
\hypersetup{
    colorlinks=true,
    linkcolor=blue,
    citecolor=blue,
    filecolor=magenta,      
    urlcolor=cyan,
    pdftitle={Overleaf Example},
    pdfpagemode=FullScreen,
    }
\makeatletter
\renewcommand*\aa@pageof{, page \thepage{} of \pageref*{LastPage}}
\makeatother

\newcommand{\lya}{Ly$\alpha$}
\newcommand{\niii}{N\,\textsc{iii}]}
\newcommand{\niv}{N\,\textsc{iv}]}
\newcommand{\hei}{He\,\textsc{i}}
\newcommand{\heii}{He\,\textsc{ii}}
\newcommand{\ciii}{C\,\textsc{iii}]}
\newcommand{\oii}{[O\,\textsc{ii}]}
\newcommand{\oiii}{[O\,\textsc{iii}]}
\newcommand{\oiiiSF}{O\,\textsc{iii}]}
\newcommand{\neiii}{[Ne\,\textsc{iii}]}
\newcommand{\niiiL}{N\,\textsc{iii}]\,$\lambda$}
\newcommand{\nivL}{N\,\textsc{iv}]\,$\lambda$}
\newcommand{\nv}{[N\,\textsc{V}]}
\newcommand{\ciiiL}{C\,\textsc{iii}]\,$\lambda$}

\newcommand{\civ}{C\,\textsc{iv}}
\newcommand{\oiiL}{[O\,\textsc{ii}]\,$\lambda\lambda$}
\newcommand{\oiiiL}{[O\,\textsc{iii}]\,$\lambda$}
\newcommand{\oiiiSFL}{O\,\textsc{iii}]\,$\lambda\lambda$}
\newcommand{\neiiiL}{[Ne\,\textsc{iii}]\,$\lambda$}
\newcommand{\hd}{H\,$\delta$}
\newcommand{\hg}{H\,$\gamma$}
\newcommand{\hb}{H\,$\beta$}
\newcommand{\mgii}{Mg\,\textsc{ii}}

\newcommand{\oiiiauL}{[O\,\textsc{iii}]\,$\lambda$}
\newcommand{\xhi}{$x_\textsc{hi}$}

\newcommand{\jwst}{\emph{JWST}}
\newcommand{\flux}{erg\,s$^{-1}$\,cm$^{-2}$}
\newcommand{\kms}{km\,s$^{-1}$\xspace}
\newcommand{\asec}{^{\prime\prime}}
\begin{document}

   \title{JADES NIRSpec Spectroscopy of GN-z11: Lyman-$\alpha$ emission and possible enhanced nitrogen abundance in a $z=10.60$ luminous galaxy}

    \titlerunning{JADES Spectroscopy of GN-z11}

\author{
Andrew J.\ Bunker
\inst{1}\fnmsep\thanks{andy.bunker@physics.ox.ac.uk}
\and
Aayush Saxena
\inst{1}\fnmsep\inst{2}
\and
Alex J.\ Cameron
\inst{1}
\and
Chris J.\ Willott
\inst{3}
\and
Emma Curtis-Lake
\inst{4}
\and
Peter Jakobsen
\inst{5}\fnmsep\inst{6}
\and
Stefano Carniani
\inst{7}
\and
Renske Smit
\inst{8}
\and
Roberto Maiolino
\inst{9}\fnmsep\inst{10}\fnmsep\inst{2}
\and
Joris Witstok
\inst{9}\fnmsep\inst{10}
\and
Mirko Curti
\inst{9}\fnmsep\inst{10}\fnmsep\inst{11}
\and
Francesco D’Eugenio
\inst{9}\fnmsep\inst{10}
\and
Gareth C. Jones
\inst{1}
\and
Pierre Ferruit
\inst{12}
\and
Santiago Arribas
\inst{13}
\and
Stephane Charlot
\inst{14}
\and
Jacopo Chevallard
\inst{1}
\and
Giovanna Giardino
\inst{15}
\and
Anna de Graaff
\inst{16}
\and
Tobias J. Looser
\inst{9}\fnmsep\inst{10}
\and
Nora L\"utzgendorf
\inst{17}
\and
Michael V. Maseda
\inst{18}
\and
Tim Rawle
\inst{17}
\and
Hans-Walter Rix
\inst{16}
\and
Bruno Rodríguez Del Pino
\inst{13}
\and
Stacey Alberts
\inst{19}
\and
Eiichi Egami
\inst{19}
\and
Daniel J. Eisenstein
\inst{20}
\and
Ryan Endsley
\inst{21}
\and
Kevin Hainline
\inst{19}
\and
Ryan Hausen
\inst{22}
\and
Benjamin D. Johnson
\inst{20}
\and
George Rieke
\inst{19}
\and
Marcia Rieke
\inst{19}
\and
Brant E. Robertson
\inst{23}
\and
Irene Shivaei
\inst{19}
\and
Daniel P. Stark
\inst{19}
\and
Fengwu Sun
\inst{19}
\and
Sandro Tacchella
\inst{9}\fnmsep\inst{10}
\and
Mengtao Tang
\inst{19}
\and
Christina C. Williams
\inst{24}\fnmsep\inst{19}
\and
Christopher N. A. Willmer
\inst{19}
\and
William M. Baker
\inst{9}\fnmsep\inst{10}
\and
Stefi Baum
\inst{25}
\and
Rachana Bhatawdekar
\inst{12}\fnmsep\inst{26}
\and
Rebecca Bowler
\inst{27}
\and
Kristan Boyett
\inst{28}\fnmsep\inst{29}
\and
Zuyi Chen
\inst{19}
\and
Chiara Circosta
\inst{12}
\and
Jakob M. Helton
\inst{19}
\and
Zhiyuan Ji
\inst{19}
\and
Jianwei Lyu
\inst{19}
\and
Erica Nelson
\inst{30}
\and
Eleonora Parlanti
\inst{7}
\and
Michele Perna
\inst{13}
\and
Lester Sandles
\inst{9}\fnmsep\inst{10}
\and
Jan Scholtz
\inst{9}\fnmsep\inst{10}
\and
Katherine A. Suess
\inst{23}\fnmsep\inst{31}
\and
Michael W. Topping
\inst{19}
\and
Hannah \"Ubler
\inst{9}\fnmsep\inst{10}
\and
Imaan E. B. Wallace
\inst{1}
\and
Lily Whitler
\inst{19}
}

\institute{
Department of Physics, University of Oxford, Denys Wilkinson Building, Keble Road, Oxford OX1 3RH, UK
\and
Department of Physics and Astronomy, University College London, Gower Street, London WC1E 6BT, UK
\and
NRC Herzberg, 5071 West Saanich Rd, Victoria, BC V9E 2E7, Canada
\and
Centre for Astrophysics Research, Department of Physics, Astronomy and Mathematics, University of Hertfordshire, Hatfield AL10 9AB, UK
\and
Cosmic Dawn Center (DAWN), Copenhagen, Denmark
\and
Niels Bohr Institute, University of Copenhagen, Jagtvej 128, DK-2200, Copenhagen, Denmark
\and
Scuola Normale Superiore, Piazza dei Cavalieri 7, I-56126 Pisa, Italy
\and
Astrophysics Research Institute, Liverpool John Moores University, 146 Brownlow Hill, Liverpool L3 5RF, UK
\and
Kavli Institute for Cosmology, University of Cambridge, Madingley Road, Cambridge CB3 0HA, UK
\and
Cavendish Laboratory, University of Cambridge, 19 JJ Thomson Avenue, Cambridge CB3 0HE, UK
\and
European Southern Observatory, Karl-Schwarzschild-Strasse 2, D-85748 Garching bei Muenchen, Germany
\and
European Space Agency (ESA), European Space Astronomy Centre (ESAC), Camino Bajo del Castillo s/n, 28692 Villanueva de la Cañada, Madrid, Spain
\and
Centro de Astrobiolog\'ia (CAB), CSIC–INTA, Cra. de Ajalvir Km.~4, 28850- Torrej'on de Ardoz, Madrid, Spain
\and
Sorbonne Universit\'e, CNRS, UMR 7095, Institut d'Astrophysique de Paris, 98 bis bd Arago, 75014 Paris, France
\and
ATG Europe for the European Space Agency, ESTEC, Noordwijk, The Netherlands
\and
Max-Planck-Institut f\"ur Astronomie, K\"onigstuhl 17, D-69117, Heidelberg, Germany
\and
European Space Agency, Space Telescope Science Institute, Baltimore, Maryland, USA
\and
Department of Astronomy, University of Wisconsin-Madison, 475 N. Charter St., Madison, WI 53706 USA
\and
Steward Observatory, University of Arizona, 933 N. Cherry Ave., Tucson, AZ 85721, USA
\and
Center for Astrophysics $|$ Harvard \& Smithsonian, 60 Garden St., Cambridge, MA 02138, USA
\and
Department of Astronomy, University of Texas, Austin, TX 78712, USA
\and
Department of Physics and Astronomy, The Johns Hopkins University, 3400 N. Charles St., Baltimore, MD 21218, USA
\and
Department of Astronomy and Astrophysics, University of California, Santa Cruz, 1156 High Street, Santa Cruz, CA 95064, USA
\and
NSF’s National Optical-Infrared Astronomy Research Laboratory, 950 North Cherry Avenue, Tucson, AZ 85719, USA
\and
Department of Physics and Astronomy, University of Manitoba, Winnipeg, MB R3T 2N2, Canada
\and
European Space Agency, ESA/ESTEC, Keplerlaan 1, 2201 AZ Noordwijk, NL
\and
Jodrell Bank Centre for Astrophysics, Department of Physics and Astronomy, School of Natural Sciences, The University of Manchester, Manchester, M13 9PL, UK
\and
School of Physics, University of Melbourne, Parkville 3010, VIC, Australia
\and
ARC Centre of Excellence for All Sky Astrophysics in 3 Dimensions (ASTRO 3D), Australia
\and
Department for Astrophysical and Planetary Science, University of Colorado, Boulder, CO 80309, USA
\and
Kavli Institute for Particle Astrophysics and Cosmology and Department of Physics, Stanford University, Stanford, CA 94305, USA
}

   \authorrunning{A.\ J.\ Bunker et al.}
   \date{}

 
  \abstract
{We present JADES JWST/NIRSpec spectroscopy of GN-z11, the most luminous candidate $z>10$ Lyman break galaxy in the GOODS-North field with $M_{UV}=-21.5$. We derive a redshift of $z=10.603$ (lower than previous determinations) based on multiple emission lines in our low and medium resolution spectra over $0.8-5.3\,\mu$m. We significantly detect the continuum and measure a blue rest-UV spectral slope of $\beta=-2.4$. Remarkably, we see spatially-extended Lyman-$\alpha$ in emission (despite the highly-neutral IGM expected at this early epoch), offset 555\,km\,s$^{-1}$ redward of the systemic redshift. From our measurements of collisionally-excited lines of both low- and high-ionization (including [O\,\textsc{ii}]\,$\lambda3727$, [Ne\,\textsc{iii}]\,$\lambda 3869$ and C\,\textsc{iii}]\,$\lambda1909$) we infer a high ionization parameter ($\log U\sim -2$). We detect the rarely-seen N\,\textsc{iv}]\,$\lambda1486$ and N\,\textsc{iii}]\,$\lambda1748$ lines in both our low and medium resolution spectra, with other high ionization lines seen in the low resolution spectrum such as He\,\textsc{ii} (blended with O\,\textsc{iii}]) and C\,\textsc{iv} (with a possible P-Cygni profile). Based on the observed rest-UV line ratios, we cannot conclusively rule out photoionization from AGN, although the high C\,\textsc{iii}]/He\,\textsc{ii} and N\,\textsc{iii}]/He\,\textsc{ii} ratios are compatible with a star-formation explanation. If the observed emission lines are powered by star formation, then the strong N\,\textsc{iii}]\,$\lambda1748$ observed may imply an unusually high $N/O$ abundance. Balmer emission lines (H$\gamma$, H$\delta$) are also detected, and if powered by star formation rather than an AGN we infer a star formation rate of $\sim 20-30\,M_{\odot}\,\textrm{yr}^{-1}$ (depending on the IMF) and low dust attenuation. Our NIRSpec spectroscopy confirms that GN-z11 is a remarkable galaxy with extreme properties seen 430\,Myr after the Big Bang.}

   \maketitle
%

\section{Introduction}
Spectroscopically confirming galaxies formed within the first few hundred million years after the Big Bang, and understanding their nature and evolution, represents one of the biggest challenges of modern astrophysics, and one of the main drivers behind the \emph{James Webb Space Telescope (JWST)}. Probing the formation of some of the very first galaxies helps establish the epoch of first light in the Universe, i.e.\ the timescales of the formation of the first stars, bringing an end to the so-called cosmic Dark Ages. Deep NIRSpec follow-up of some of the highest redshift galaxy candidates has already yielded spectroscopic confirmations via clear detection of the Lyman break in four galaxies at $z>10$ \citep{Curtis-Lake2022, Robertson2022}. 

The onset of the first star formation began the process of reionizing the intergalactic medium (IGM), although the exact details of this reionization are still uncertain. Observations of Lyman-$\alpha$ emission and absorption provide strong constraints on how and when the diffuse gas in the IGM transitions from neutral to ionized (see \citealt{Robertson22a} for a review). An important observation is the decrease in Lyman-$\alpha$ emission line equivalent width with increasing redshift above $z=6$ (\citealt{Stark2010,Schenker2014,Caruana2014,Jung2020}), consistent with stronger absorption from an increasingly neutral IGM. However, this picture was largely based on moderate-luminosity galaxies. Luminous galaxies at $7.5<z<9$ often show Lyman-$\alpha$ emission (\citealt{Zitrin2015,Oesch2015,Stark2017,Larson2023,Finkelstein2013,Roberts-Borsani2016,Jung2019,Song2106}), at a redshift where quasar damping wing studies suggest the IGM is significantly neutral (\xhi \,$\sim 0.5$; \citealt{Greig2017,Davies2018,Wang2020}). This indicates that around luminous (and potentially massive) galaxies, Lyman-$\alpha$ escapes more easily, perhaps as a result of ionized bubbles that grew early in overdense regions (\citealt{Endsley2022,Jung2022}; cf. \citealt{Saxena2023}). Another possible Lyman-$\alpha$ escape mechanism is through resonant scattering, with high velocity neutral gas (perhaps associated with outflows) redshifting the photons to lower frequencies at which they are no longer absorbed by the intervening neutral IGM (\citealt{Dijkstra2014,Mason2018}). Deep spectroscopy with \jwst\ is  vastly increasing our knowledge in this area, both by detecting rest-frame optical lines of known Lyman-$\alpha$ emitters to derive systemic redshifts and details of nebular physical conditions, and by detecting Lyman-$\alpha$ further into the reionization epoch than has been possible from the ground \citep[e.g.][]{Tang2023}.

Before the launch of \jwst, the most distant galaxy with a tentative but plausible spectroscopic redshift was GN-z11 \citep{Oesch2016}.
This was first selected as a likely high redshift Lyman-break candidate through multi-colour imaging with \emph{HST} \citep{Bouwens2010, Oesch2015}, and subsequent \emph{HST/WFC3} slitless grism spectroscopy revealed a possible Lyman break in the continuum \citep{Oesch2016}, yielding a redshift of $z_{\textrm{grism}}=11.09$.  With an apparent $H_{\textrm{160}}$ magnitude of $26.0\pm0.1$, GN-z11 is remarkably bright, up to 3 times more luminous than the characteristic rest-UV luminosity ($L_\star$) measured from luminosity functions at $z\sim6-8$ \citep[e.g.][]{Finkelstein2015, Bouwens2015}. Using \emph{Spitzer}/IRAC fluxes, \citet{Oesch2016} estimated its stellar mass to be $M_\star = 10^9\,M_\odot$, indicating a rapid build up of stellar mass in the very early Universe.

Through ground-based near-infrared spectroscopy using MOSFIRE on the Keck Telescope, the redshift of GN-z11 was further refined by \citet{Jiang2021} via the possible detection of the [\ciii$\lambda1907+$\,\ciii$\lambda 1909$ doublet, yielding a redshift of $z=10.957$. If real, the intense \ciii\ emission line might originate partly due to an active galactic nucleus (AGN) hosted by the galaxy, or due to rapid carbon enrichment \citep{Jiang2021}.

Given the unique nature of this source and the low signal-to-noise ratio ($S/N$) of existing continuum break and emission line detections of this galaxy, NIRSpec on \jwst\ now offers a chance to confirm its true distance and nature through high $S/N$ detection of multiple rest-UV and optical emission lines as well as its bright continuum. Several diagnostics relying on the ratios and strengths of rest-UV and optical emission lines can help differentiate between photoionization due to AGN or star-formation alone, and further help characterize the ionization conditions in the interstellar medium (ISM) of this remarkably luminous distant galaxy. 
 
In this paper, we report an unambiguous spectroscopic redshift of $z=10.6034$ for GN-z11 using deep NIRSpec observations in the GOODS-North field via the robust detection of several emission lines including \nivL\,$1486$, \niii$\lambda \lambda1747,1749$, [\ciii$\lambda1907+$\,\ciii$\lambda 1909$, \oii\,$\lambda\lambda 3726, 3729$, \neiii\,$\lambda\lambda 3869, 3967$, \hd\ and \hg . Although the measured redshift is lower than previously reported in other work, this still places GN-z11 as comfortably the most luminous source currently confirmed at $z>10$. This galaxy is given the designation JADES-GN-z10-0 in the JADES spectroscopic database, but for the remainder of this paper we use the more familiar name GN-z11. The photometric properties from our {\em JWST}/NIRCam imaging are reported in a companion paper \citep{Tacchella2023}.

The layout of this paper is as follows. In Section~\ref{sec:observations} we describe our JWST/NIRSpec observations of GN-z11 and the data reduction strategies adopted. In Section~\ref{sec:results} we present the 1D and 2D spectra of GN-z11, emission line measurements and discuss the inferred physical properties. In Section~\ref{sec:conclusions} we conclude our findings. Throughout this work, we assume the Planck 2018 cosmology \citep{Planck18} and the AB magnitude system \citep{Oke1983}.

 \section{Observations}
\label{sec:observations}

The NIRSpec observations of GN-z11 were taken as part of the {\em JWST} Advanced Deep Extragalactic Survey (JADES), a collaboration between the Instrument Science Teams of NIRSpec and NIRCam to study galaxy evolution out to high redshift through imaging and spectroscopy in the two GOODS fields. The Guaranteed Time Observations presented here are part of program ID 1181 (P.I.: D.\ Eisenstein), with spectroscopic targets based on pre-\emph{JWST} data, largely \emph{HST} imaging.

Our observations of GN-z11 were taken on UT\ 5 \& 7 February 2023, using NIRSpec \citep{jak22} in its microshutter array (MSA) mode \citep{ferr22}. The MSA comprises four arrays of $365\times 171$ independently-operable shutters, each covering $98''\times 91''$ on the sky. GN-z11 was targeted in four independent MSA configurations. Each configuration acquired 3100\,s of integration in each of the medium-resolution G140M/F070LP, G235M/F170LP, and G395M/F290LP grating/filter combinations (with resolving power $R\approx 1000$ and combined spectral coverage over $0.7-5.3\,\mu$m) and 6200\,s in the low-resolution PRISM/CLEAR mode (with $R\sim 100$ and continuous coverage $0.7-5.3\,\mu$m). Targets were assigned three shutter slitlets, with the targets nodded into each of the three shutters during the observing sequence to facilitate background subtraction.  As GN-z11 was one of our highest priority targets, we ensured that its spectra did not overlap with other targets, even for the gratings (where the spectra are more extended on the detector than the low-dispersion prism). Individual integrations used the NRSIRS2 readout mode with 14 groups (1035\,s each) to limit correlated readout noise. In total our integration time was 3.45\,hours in each of the three gratings, and 6.9\,hours in the prism.

\begin{figure}
    \centering
    \includegraphics[width=\linewidth]{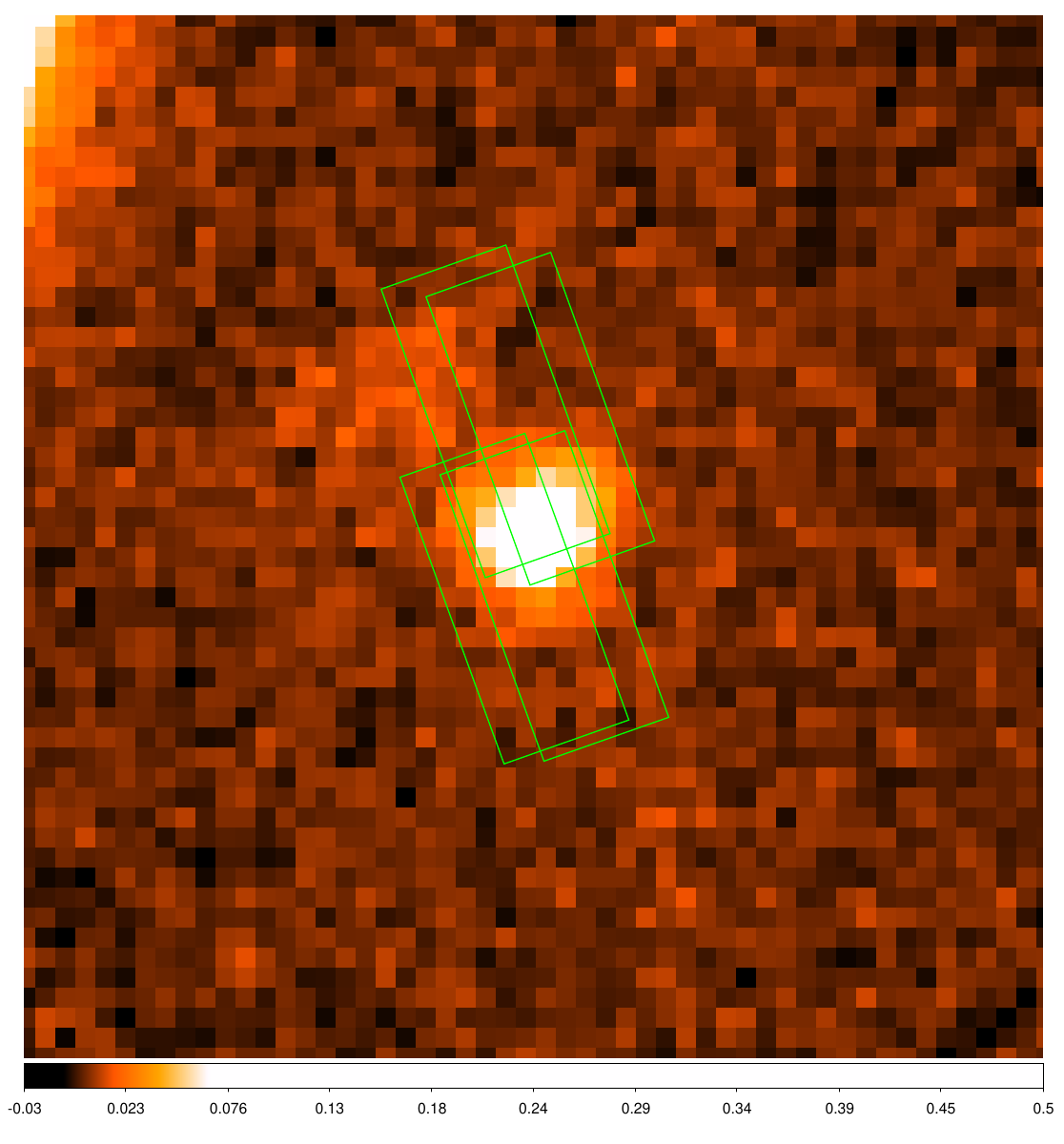}
    \caption{The NIRCam F200W image of GN-z11 (see \citealt{Tacchella2023}) with the NIRSpec microshutters overlaid for the four different pointings. The green rectangles denote the illuminated region of each microshutter ($0\farcs2 \times 0\farcs 46$). North is up and East to the left, and the image is 3\,arcsec on the side. The flux density units on the colour bar are MJy/sr.}
    \label{fig:slit_overlay}
\end{figure}

These observations were processed with algorithms developed by the ESA NIRSpec Science Operations Team and the NIRSpec GTO Team. We refer the reader to \citet{Cameron2023} for more details of the processing. We note that for the  G140M/F070LP grating/filter combination we extended the calibration of the spectrum up to 1.84 $\mu$m, taking into account the transmission filter throughput beyond the nominal wavelength range of this configuration (0.70\,$\mu$m\,–\,1.27\,$\mu$m). Since GN-z11 is at $z>10$ with no flux at wavelengths below Lyman-$\alpha$, there is no second order light to overlap with the extended wavelength range of  1.27\,$\mu$m\,–\,1.84\,$\mu$m. Wavelength-dependent path-loss corrections were made based on the object position within the shutter and modelling the galaxy as a point-like source. GN-z11 is very compact, so this is a good approximation. In all four of the configurations, GN-z11 was located not more than 40\% of the illuminated slit width or slit height from the centre along either axis of the $0\farcs20\times 0\farcs46$ slitlet, and Figure~\ref{fig:slit_overlay} shows the locations of the open areas of the microshutters overlaid on the NIRCam F200W image of GN-z11. Individual calibrated one-dimensional (1D) and 2D spectra  were combined excluding bad pixels  by using an an iterative sigma clipping algorithm at the 3$\sigma$ level. The wavelength calibration takes into account the position of the galaxy within the slit at each pointing. Before the combination process a spatial median of each exposure is subtracted to remove any residual background. 1D spectral extractions were made from the rectified 2D spectra using box extractions of height 3 and 5 pixels ($0\farcs 3$ and $0\farcs5$, respectively).

\section{Results}
\label{sec:results}
\begin{figure*}
   \centering
   \includegraphics[width=\linewidth]{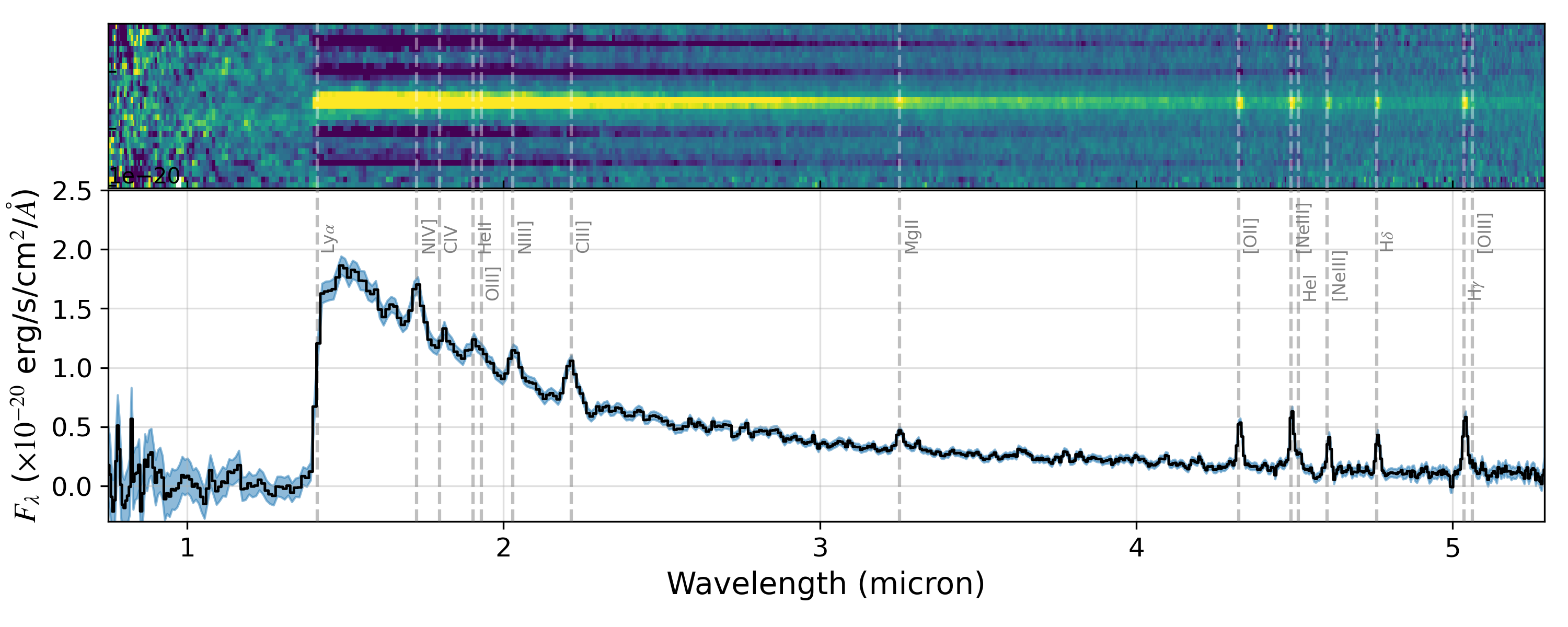}
    \caption{2D (top) and 1D (bottom) spectra of GN-z11 using PRISM/CLEAR configuration of NIRSpec. The 1D spectrum has been extracted using a 3 pixel wide aperture that leads to improved S/N in this highly compact object. Prominent emission lines present in the spectra are marked. The signal to noise ratio (SNR) of the continuum is high and the emission lines are clearly seen in both the 1D and 2D spectra.}
    \label{fig:spectrum}
\end{figure*}

\begin{figure*}
    \centering
    \includegraphics[width=\linewidth]{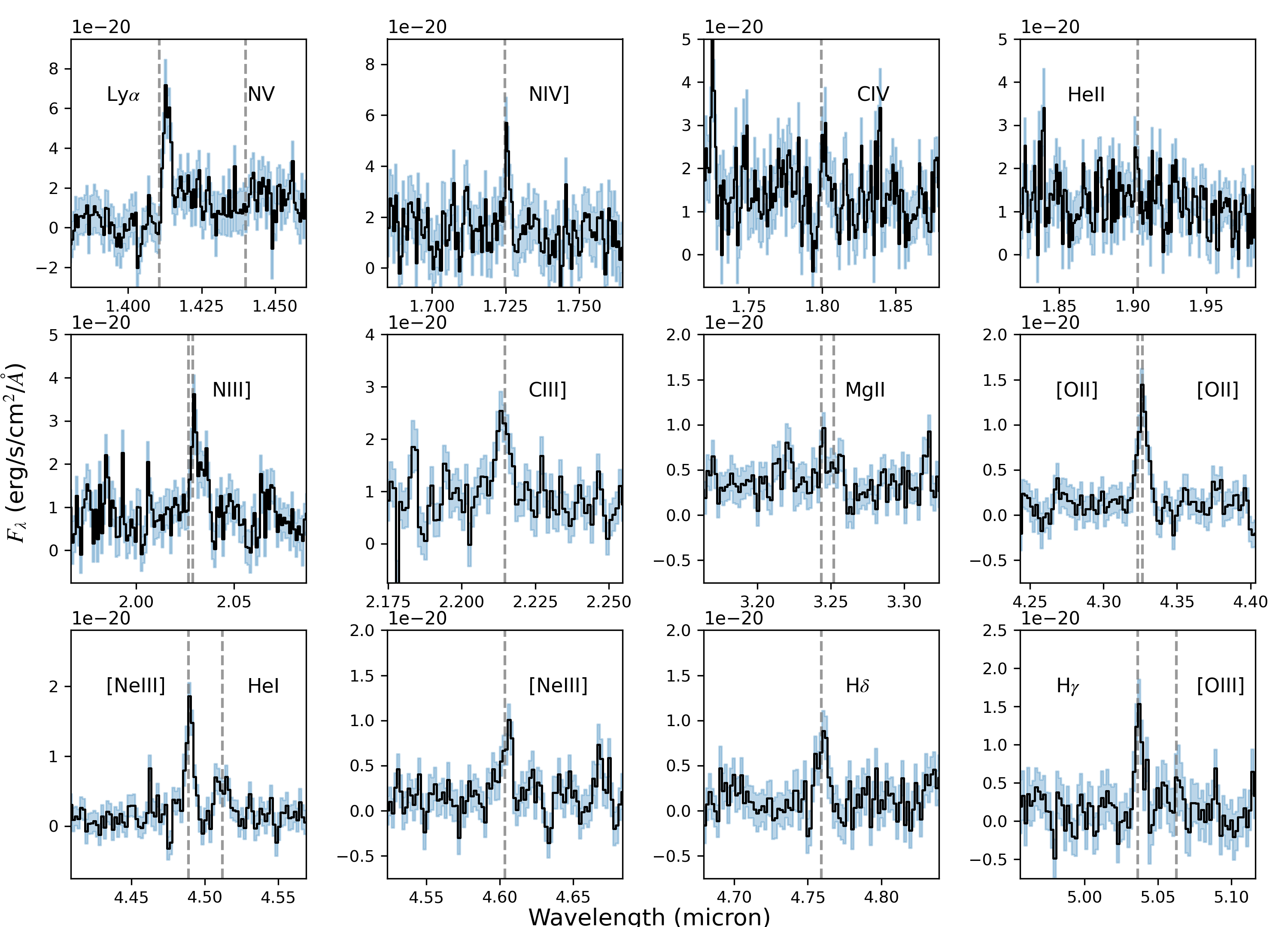}
    \caption{Gallery of the most prominent emission lines seen in the spectrum GN-z11 from the Medium resolution (R1000) gratings using a 3 pixel 1D spectral extraction.}
    \label{fig:emlines}
\end{figure*}

Our NIRSpec spectra of GN-z11 show well-detected continuum in the prism (Figure \ref{fig:spectrum}) where we have $S/N>20$ per spectral resolution element at wavelength above Lyman-$\alpha$ and out to $\sim3\,\mu$m. We also see the continuum at lower $S/N$ in the medium-resolution gratings (Figure \ref{fig:emlines}, Figure \ref{fig:gratings}). A strong spectral break at Lyman-$\alpha$ is observed, with no significant flux at shorter wavelengths. We have robust detections of several emission lines, most of which are seen in both the prism (Figure \ref{fig:spectrum}) and grating (Figure \ref{fig:emlines}) spectra. We also see evidence of interstellar absorption lines, but in this paper we focus on the emission lines properties.

In the subsections below, we use the spectrum of GN-z11 to infer physical properties.
As well as using empirical diagnostics from the emission line fluxes and ratios, we also use the \textsc{beagle} Bayesian SED fitting code \citep{Chevallard2016} on our full prism spectrum, the exact details and results are presented in Table \ref{tab:beagle} and in Appendix \ref{sec:beaglefit}.

\subsection{Emission lines and Redshift Determination}
\label{sec:redshift}

The full list of detected lines is given in Table \ref{tab:lines}. Line wavelengths are measured from the grating spectra because they have higher resolution resulting in less blending and more accurate line centroids. Line fluxes are measured from both the prism and gratings, and we discuss the relative flux calibration between the two in Appendix~\ref{sec:compareNIRCam}. We perform a cubic-spline fit to the continuum in the prism (excluding the emission lines from the fit), and use this to subtract the continuum level in both the prism and the grating spectra (since the continua in the grating spectra have low signal-to-noise ratio). We then fit each emission line with a single Gaussian model, except for blended lines (noted in Table \ref{tab:lines}) where we measure the total flux of the complex.  The uncertainty is computed taking into account the Poisson counting statistics and the readout noise. This is preferred to taking the intra-pixel standard deviation in counts in the spectrum, which would underestimate the true noise since the pixels are sub-sampled and interpolated from their native size by the data reduction pipleline and hence the noise is correlated.
One emission line, \nivL$1486$, falls in the spectral coverage of both the G140M and G235M gratings, and we report both measurements in Table~\ref{tab:lines}. The wavelength and flux of this line are consistent between the gratings within the uncertainties.

In determining the redshift from the vacuum rest-frame wavelengths, we exclude Lyman-$\alpha$ (which has a velocity offset, see Section~\ref{subsec:lya}), \mgii\ (which is only significantly detected in the low-resolution prism) and \hei\,$ \lambda 3889$ (which is blended with Balmer-8). A weighted fit of the remaining 8 well-detected emission lines ($S/N>5$) from the grating spectra give a redshift $z=10.6034 \pm 0.0013$, where we assume a  [C\,\textsc{iii}]\,$\lambda1907$ \,/\, \ciii$\lambda 1909$ doublet ratio of $1.5$ for these spectrally-unresolved lines.

The redshift we measure is considerably lower than the previously reported redshift values of $z = 11.09 ^{+0.08} _{-0.012}$ from \emph{HST} grism \citep{Oesch2016} and $z = 10.957 \pm 0.001$ from Keck MOSFIRE \citep{Jiang2021}. The 2D \emph{HST} grism observation shows flux down to the wavelength we measure for the Lyman break ($1.41\,\mu$m), but due to noise fluctuations their fitted model break was at a longer wavelength of $1.47\,\mu$m. The Keck MOSFIRE redshift was based on possibles detections of the [C\,\textsc{iii}]\,$\lambda1907$ and \ciii$\lambda 1909$ lines at $2.2797\,\mu$m and $2.282\,\mu$m respectively, at $2.6\,\sigma$ and $5.3\,\sigma$. We do not find any significant emission lines at these observed wavelengths in our data, where they would have been detected at $20\,\sigma$ and $40\,\sigma$ for the line fluxes quoted in \citet{Jiang2021}. Instead, we do detect \ciii\ but at a shorter wavelength consistent with our measured $z=10.603$.

\begin{table*}
    \centering
    \caption{Emission line fluxes (in units of $\times10^{-19}$\,\flux) detected in the prism ($R\sim100$) and grating ($R\sim1000$) spectra measured from both the 3 pixel and the 5 pixel extraction.}
    \begin{tabular}{l c c c c c c}
    \hline
    Emission line  & $\lambda_\mathrm{obs}$ (\AA )  & $F_{\textrm{R100}}^{\textrm{3pix}}$  &  $F_{\textrm{R100}}^{\textrm{5pix}}$  &  $F_{\textrm{R1000}}^{\textrm{3pix}}$  & $F_{\textrm{R1000}}^{\textrm{5pix}}$  & EW$_0$ (\AA) \\
    \\
    \hline
    \\
    \lya\ & $14132.0\pm 1.0$ & $-$ & $-$ & $15.1\pm 1.5$ & $23.0\pm 1.9$  & $18.0 \pm 2.0^g$ \\
\nv\ $\lambda\lambda1238,1242$ & (undetected) & (blended Ly$\alpha$) & $-$ & $<4.2~(3\sigma)$ & $<5.4~(3\sigma)$ & $<5.0$ \\
    \nivL$1486$ (G140M) & $17254.8\pm 1.3$ & $13.6\pm 1.7$ & $12.3\pm 2.1$  & $7.9\pm 1.4$ &  $8.8\pm 1.8$  & $5.1 \pm 0.9$ \\
    \nivL$1486$ (G235M) & $17253.3\pm 1.5$ & $13.6\pm 1.7 $ & $12.3\pm 2.1$ & $10.2\pm 1.4$ &  $13.1\pm 2.2$ & $6.6 \pm 0.9$ \\
        \civ\,$\lambda \lambda 1548,1550$ (em) & $18012.3\pm 5.7$  & (P-Cygni-like) & $-$ & $4.8\pm 1.6$ & $4.6\pm 2.0$ & $3.5\pm 1.2$ \\
    \civ\,$\lambda \lambda 1548,1550$ (abs) & $17943.6\pm 3.6$  & (P-Cygni-like) & $-$ & $-7.5\pm 1.6$ & $-4.8\pm 2.0$ & $-5.3\pm 1.1$  \\
    \heii\,$\lambda1640$ & $19021.5\pm 3.3$ & $(11.1\pm 1.3)^a$ & $(14.1\pm 1.6)^a$
 & $4.2\pm 1.3$ & $4.7\pm 1.7$ & $3.5\pm 1.1$ \\
\oiiiSFL\,$1660,1666$ & $19288.2\pm 5.3$ &  $(11.1\pm 1.3)^a$ & $(14.1\pm 1.6)^a$ & $4.2\pm 1.5$ & $5.7\pm 2.0$ & $3.6\pm 1.3$\\
    \niii\,$\lambda \lambda 1749-1753^b$ & $20297.2\pm 1.5^b$ & $10.7\pm 1.2^b$ & $11.9\pm 1.5^b$  & $13.8\pm 1.6^b$ & $14.4\pm 1.5^b$  & $13.4 \pm 1.6$ \\
\ciii\,$\lambda \lambda 1907,1909$ & $22139.2\pm 2.0$ & $13.3\pm 1.1$ & $15.0\pm 1.3$  & $10.5\pm 0.9$ & $12.2\pm 1.2$  & $12.5 \pm 1.1$  \\
    \mgii\,$\lambda \lambda 2795,2802^c$ & $32447.6\pm 4.3^c$ & $4.1\pm 0.5^c$ & $5.4\pm 0.6^c$ & $4.3\pm 0.9^c$ & $4.3\pm 0.9^c$ & $11.7\pm 2.5$ \\
    \oii\,$\lambda \lambda 3726,3729$ & $43265.8\pm 2.8$ & $7.1\pm 0.4$ & $8.7\pm 0.6$ & $8.9\pm 0.8$ & $10.0\pm 0.9$  & $45.3 \pm 4.1$ \\
    \neiii\,$\lambda 3869$ & $44896.8\pm 1.9$ & $(12.0\pm 0.6)^d$ & $(10.1\pm 0.5)^d$  & $10.0\pm 0.8$ & $11.0\pm 1.0$  & $57.9 \pm 4.6$ \\
    \hei\,$ \lambda 3889\,+$\,H8 & $45119.3\pm 7.9$ & $(12.0\pm 0.6)^d$ & $(10.1\pm 0.5)^d$  & $5.7\pm 0.9$ & $6.6\pm 0.9$  & $33.7 \pm 5.3$ \\
    \neiii\,$\lambda 3967$ + H$\epsilon$ & $46052.3\pm 3.7$ & $4.1\pm 0.5$ & $4.7\pm 0.5$  & $5.9\pm 0.7$ & $6.3\pm 0.8$  & $38.0 \pm 4.5$ \\
    \hd\ & $47600.9\pm 4.9$  & $5.1\pm 0.4$ & $6.3\pm 0.6$  & $6.0\pm 0.9$ & $8.2\pm 1.1$  & $44.4 \pm 6.7$ \\
    \hg\ & $50370.1\pm 3.5$ & $9.6\pm 0.5$ & $12.0\pm 0.6$ & $8.8\pm 1.2$ & 
    $10.0\pm 1.5$  & $73.3 \pm 10.0$ \\
    \oiiiauL$4363$ & $50639.1\pm 10.7$ & $1.6\pm 0.6^e$ & $2.3 \pm 0.7^e$ & $2.6\pm 1.4$ & $<4.5~(3\sigma)$ & $<40~(3\sigma)$\\
\\
    \hline
    \end{tabular}
    
    Emission line fluxes (in units of $\times10^{-19}$\,\flux) detected in the prism ($R\sim100$) and grating ($R\sim1000$) spectra measured from both the 3 pixel and the 5 pixel extraction. The rest-frame equivalent widths given (determined from the 3 pixel extraction) are derived using line fluxes measured from the medium resolution gratings, with the continuum measured from the lower resolution spectrum with a higher $S/N$.  \nivL$1486$ is well detected in two gratings, and both measurements are reported. \civ\,$\lambda \lambda 1548,1550$ exhibits a P-Cygni profile, and we report the grating fluxes separately for the emission and absorption components; this doublet would be self-absorbed at the prism resolution.
    
$^a$ \heii\,$\lambda1640$ is blended with \oiiiSFL $1660,1666$ at the prism resolution, and the total flux is reported. These lines are resolved in the grating but only marginally detected ($\approx 3\sigma$).

$^b$ the wavelength for the strongest component of the multiplet is reported (a blend of \nivL 1748.64 \& 1749.6), but the flux is summed over all the components in the grating rather than fitting a single Gaussian

$^c$ for the MgII doublet, the wavelength  of the first 2795\AA\ line is reported from the grating spectrum, but the total flux of this complex is reported.

$^d$ [NeIII]3869 and HeI 3889 are blended in the prism, and the total flux of the complex is reported.

$^e$ \oiiiauL$4363$ is partially-blended with H$\gamma$ at the prism resolution, and undetected in the grating.

$^f$ the \lya\ equivalent width uses the flux in the 5 pixel extraction of the grating spectrum, as this line is spatially extended
    \label{tab:lines}
\end{table*}

\subsection{Is GN-z11 an AGN?}
\label{subsec:agn}

GN-z11 has a compact morphology and the continuum spatial extent in our NIRSpec 2D spectroscopy is barely resolved. In a companion paper, \citet{Tacchella2023} analyze JADES NIRCam imaging data and derive the best size constraint so far, finding an intrinsic half-light radius of only $0.016\pm0.005\asec$ ($64\pm20$\,pc). The possibility of a significant point source contribution to the total flux leaves open the question of whether some of the light originates from an AGN. Our data do contain several high ionization lines and we wish to explore the excitation mechanism.

We have detected a large number of emission lines of varying ionization potential in GN-z11. In particular the 
\nivL1486 line (ionization potential $E>47.5$\,eV) is often a signature of an AGN, although it has been seen in some high-redshift star forming galaxies (e.g., \citealt{Fosbury2003} and \citealt{McGreer2018}) and it is detected in 6 of 44 galaxies in the low-redshift CLASSY survey \citep{Mingozzi2022}. \nivL1486 or \niiiL1748 is seen in 1\% of SDSS quasars at $1.7<z<4.0$ \citep{jiang2008}.  \cite{Vanzella2010} detect \nivL1486  in an object at $z=5.6$  which may be an AGN, although \cite{Raiter2010} are able to model this spectrum with stars. For GN-z11, the higher ionization Nitrogen line N\,\textsc{v} ($E > 77.5$\,eV), which is a clear signature of AGN activity, is undetected in the G140M grating (where the $3\,\sigma$ sensitivity is $4\times 10^{-19}$\,erg\,cm$^{-2}$\,s$^{-1}$), and is blended with Lyman-$\alpha$ continuum break at the prism resolution.

We note that in the prism ($R\sim100$) spectrum we see prominent emission features arising from blended \heii\ and \oiii\,$\lambda \lambda1660,1666$ lines, as well as a P-Cygni type feature from \civ\ (see Figure~\ref{fig:emlines}), with redshifted emission and blueshifted absorption relative to the systemic redshift. civ\ emission with a P-Cygni-like profile is also detected in the grating, along with a low $S/N$ tentative detection ($3\,\sigma$) of of \heii\  and \oiii\,$\lambda \lambda1660,1666$.

\begin{figure}
    \centering
    \includegraphics[width=\linewidth]{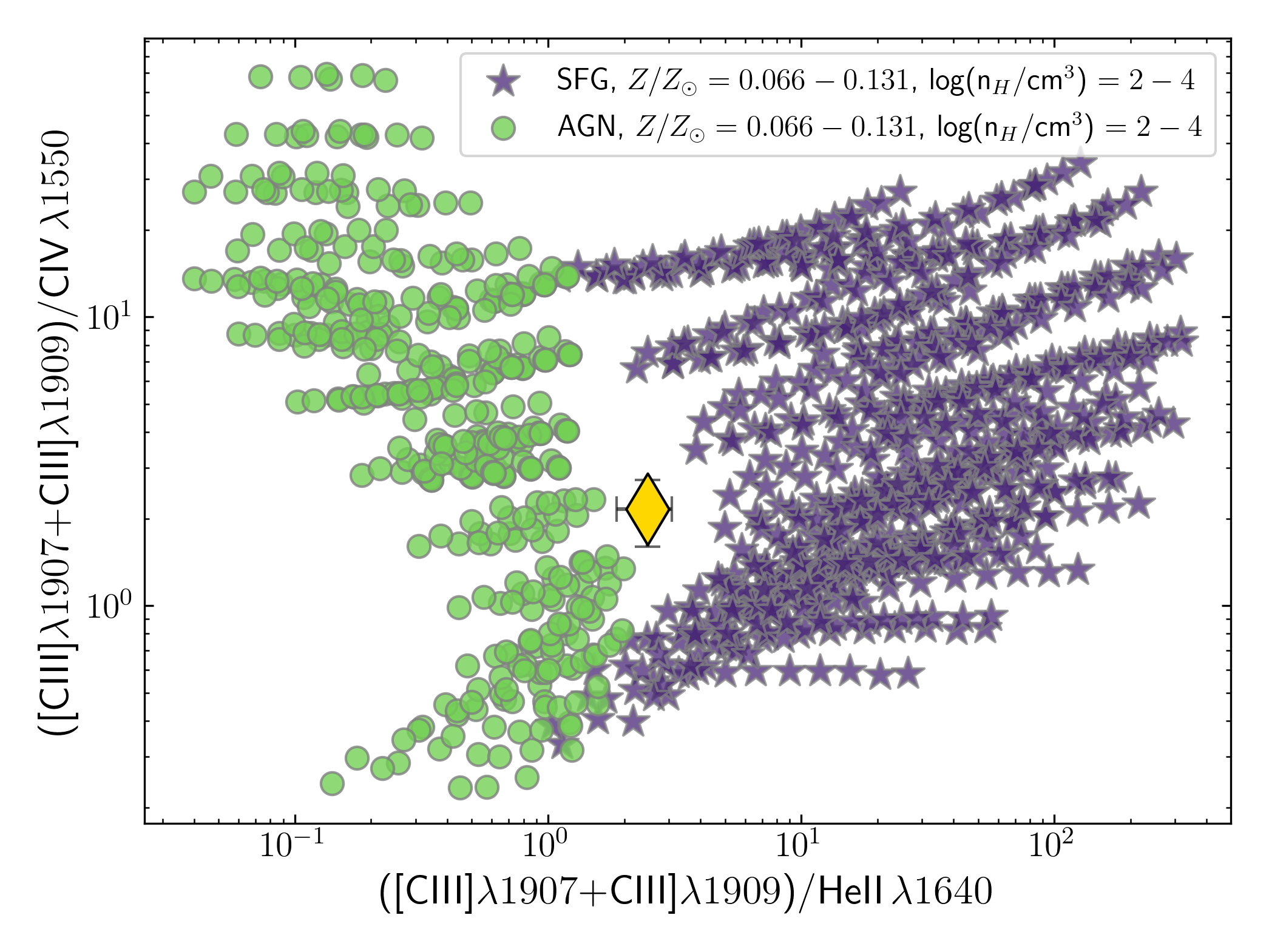}
    \caption{Measured \ciii/\heii\ vs \ciii/\civ\ ratios for GN-z11 shown along with predictions from photoionization due to AGN (circles) and star-formation (stars) from \citet{Feltre2016} and \citet{Gutkin2016} considering in the range of $Z/Z_\odot = 0.066 - 0.131$ and gas densities in the range $\log(n_H)/\textrm{cm}^{-3} = 2 - 4$. Based on the observed line ratios neither photoionization from AGN or star-formation alone can conclusively explain the observations, placing GNz11 right between the model predictions from the two.}
    \label{fig:uv}
\end{figure}

The reliable detection of \ciii\ and \niii\ lines in the grating spectra, however, enables us to investigate rest-UV line ratios that can be compared with predictions from photoionization models to differentiate between an AGN or a star-formation origin \citep[e.g.][]{Feltre2016}. In Figure~\ref{fig:uv} we plot the line ratios \ciiiL$1909$/\heii\,$\lambda1640$ versus \ciiiL$1909$/\civ\,$\lambda \lambda 1548,1550$, along with predictions from photoionization models of \citet{Feltre2016} for type 2 AGN, and \citet{Gutkin2016} for star formation. We consider a density range of  $\log(n_H/\textrm{cm}^{-3})$ range of $2-4$, and metallicities in the range $Z=0.001 - 0.002$, corresponding to $Z/Z_\odot = 0.066 - 0.131$ (based on $Z_\odot = 0.0152$ assumed by \citealt{Feltre2016}), which is consistent with the gas-phase metallicity inferred from SED fitting using \textsc{beagle} (Section \ref{sec:beaglefit}), and the estimates from emission line ratios in Section~\ref{subsec:ism}.

We find that neither AGN nor SFG model predictions are able to conclusively explain the observed line ratios in GNz11. As previously noted, there is \civ\ absorption visible in the spectrum, blueshifted from the systemic redshift. However, it is unclear how much of the nebular component of the \civ\ line flux is being attenuated by this blueshifted absorption as there is not enough S/N in the grating to disentangle the nebular and stellar components of \civ. If a significant amount of nebular \civ\ emission is also being absorbed, then the data point would move downwards on the $y$-axis in Figure \ref{fig:uv}.

Additionally, using the \ciii\ and \heii\ based diagnostics from \citet{Nakajima2018}, we find that the observed strength of \ciii\ emission and the \ciii/\heii\ ratio once again lie between photoionization model predictions due to type-2 AGN and star-formation. \citet{Nakajima2018} found a parameter space in their diagnostic plots where both AGN and star-forming models could overlap due to low metallicities and high C/O ratios, which is where the measurements from GN-z11 suggest it could lie. We note that when considering the photoionization models of \citet{Nakajima2022b}, the limit on \ciii/\civ\ we derive is compatible with the envelope of expectations from AGN over a range of metallicities.

A ratio of \niii/\heii\,$\approx3.3$ is consistent with photoionization due to star-formation \citep[e.g.][]{Hirschmann2019}, and interestingly, no type 2 AGN scenario in the models of \citet{Hirschmann2019} predicts a \niii/\heii\ ratio greater than~1. Composite models containing contribution from both AGN and star-formation can achieve \niii/\heii\ ratios $\sim1$, but only star-formation is favoured at ratios $>1$.

Overall, we find that the \ciii\ and \niii\ emission and their ratios with respect to \heii\ and \civ\ do not obviously favour photoionization due to AGN. However, the presence of other rare lines (e.g. \niv) that have previously been observed in the spectra of AGN makes ruling out the presence of an AGN less obvious \citep[see][for example]{Ubler2023}. Given the expected extreme nature of GN-z11, together with a lack of any observational insights into the expected spectroscopic properties of AGN at $z>10$, we are unable to draw definitive conclusions about the dominant source of photoionization in GN-z11.

Finally, we note that the grating spectra do not show obvious evidence for the presence of a broad component of permitted lines (see Figure~\ref{fig:emlines}), which would be ascribed to the Broad Line Region (BLR) of an AGN. This is not necessarily conclusive proof against the AGN scenario, as the BLR is often obscured along our line of sight in most AGN, however it is another element consistent with the lack of dominant contribution from an AGN.

\subsection{\texorpdfstring{Lyman-$\alpha$}{Ly-a} Emission}
\label{subsec:lya}

The prism spectrum shows a near-total Gunn-Peterson trough at wavelengths below Lyman-$\alpha$, consistent with a highly neutral intervening IGM \citep{Gunn1965}. Although the spectral break is fairly sharp in wavelength at the low dispersion of the prism, we do see some evidence of a damping wing absorption. 

\begin{figure}
    \centering
    \includegraphics[width=\linewidth]{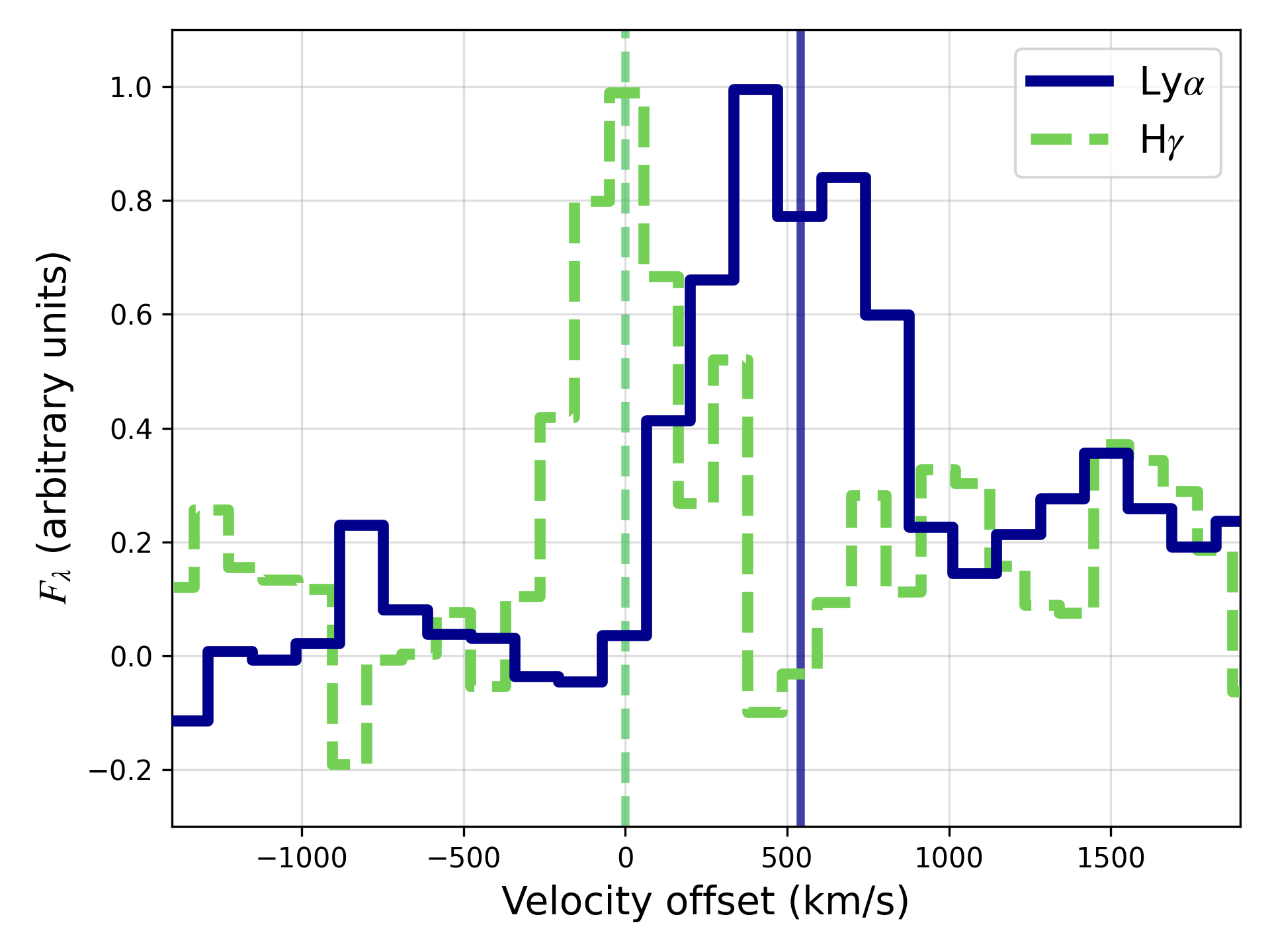}
    \caption{Velocity offset of the \lya\ emission line (blue solid line) compared with the \hg\ line (green dashed line). The \lya\ line is redshifted by $555$\,\kms\ compared to the redshift derived from other emission lines in the spectrum.}
    \label{fig:lya_veloffset}
\end{figure}

In spectra from the bluest G140M grating, an emission line is seen at 14132\,\AA, close to the sharp Lyman break observed with the prism. Taking the systemic redshift of GN-z11 to be $z=10.6034$ (see Section~\ref{sec:redshift}), the rest-frame wavelength is 1217.92\,\AA , consistent with being Lyman-$\alpha$ in emission, but with the line centroid redshifted by $555\pm 32$\,\kms\ (see Figure~\ref{fig:lya_veloffset}). This emission line is seen in all 12 of the individual G140M grating exposures (in each of the 3 nod positions in the 4 MSA configurations), and is $>10\,\sigma$ in the combined grating spectrum. 

\begin{figure}
    \centering
    \includegraphics[width=\linewidth]{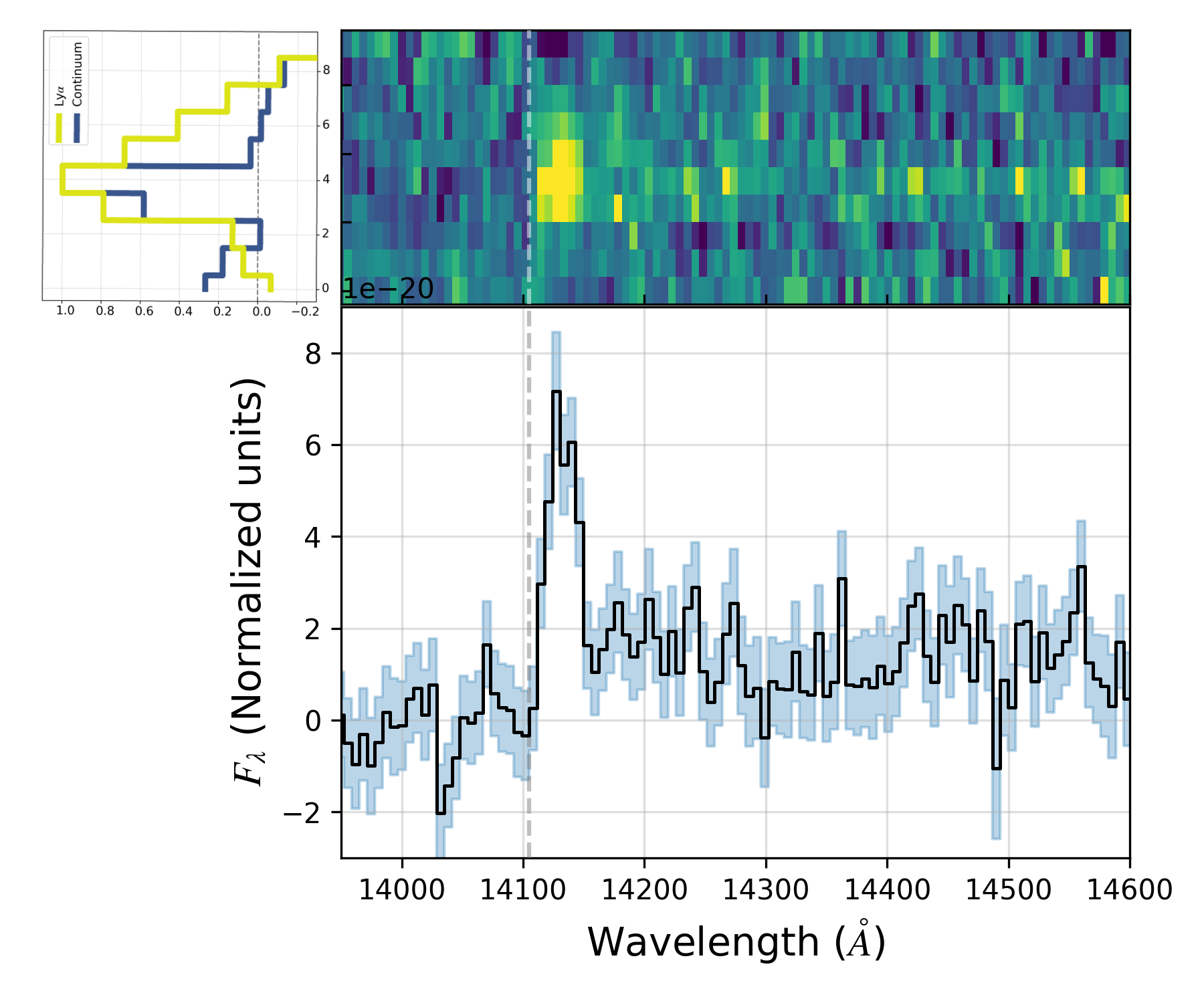}
    \caption{Zoom in on the \lya\ emission line in the G140M 1D (lower) and 2D (upper) spectra. The grey dashed line shows the systemic wavelength of the \lya\ transition. The histogram (top-left) of the \lya\ spatial profile (yellow) and that of the continuum (blue), shows the \lya\ emission from GN-z11 is more extended towards the south-west (up in the MSA shutter in this view).}
    \label{fig:Lya_zoom}
\end{figure}

Figure \ref{fig:Lya_zoom} shows a close-up of the G140M spectral region around Lyman-$\alpha$. There is zero transmitted flux shortward of the systemic Lyman-$\alpha$ wavelength. Any such flux would require an ionized bubble around the galaxy, but the lack of such flux rules out an optically-thin H\,\textsc{II} region around GN-z11 \citep[e.g.][]{Mason2020}. The redshifted line is well approximated by a Gaussian, without significant asymmetry. We measure a FWHM of $566\pm 61$\,\kms, which is extended beyond the instrumental line spread function of $\approx 200$\,\kms\ (de Graaff et al. in prep) for a compact source in the G140M grating at this wavelength. Removing the line spread function in quadrature suggests an intrinsic velocity spread of $\delta v_{\textrm{FWHM}}=530\pm 65$\,\kms.

In the 2D spectrum of Figure \ref{fig:Lya_zoom} it is apparent that the Lyman-$\alpha$ emission is more spatially extended than the continuum. Whilst the continuum flux is largely contained within 2 pixels (0.2$\asec$), as expected based on the small size measured in our NIRCam imaging presented in \citet{Tacchella2023}, the Lyman-$\alpha$ emission extends further to the south-west. The Lyman-$\alpha$ extension beyond the continuum is at least 2 pixels, corresponding to an extra 0.8\,kpc. We note the Lyman-$\alpha$ could extend further since the MSA shutters are only 5 pixels high, so beyond this region there is self-subtraction, but a visual check of the 2D spectrum without background subtraction did not show Lyman-$\alpha$ in neighbouring shutters. A similar check on the extent of other well-detected lines in the grating spectra (${\textrm \oii}$ and H$\gamma$) shows that these lines have the same spatial profile as their nearby continuum.

The fact that Lyman-$\alpha$ is spatially extended is a remarkable result, which may be suggestive of a Lyman-$\alpha$ halo. The presence of such haloes around individual star-forming galaxies has been reported at lower redshifts \citep[e.g.][]{Rauch2008, Wisotzki2016, Leclercq2017, Kusakabe2022} and we may be seeing the gas in the circum-galactic medium (CGM), from Lyman-$\alpha$ fluorescence or shock heating.

Using a 3 pixel ($0.3\,\asec$) extraction aperture, the measured Lyman-$\alpha$ flux is $(1.51\pm 0.15)\times 10^{-18}$\,\flux , with a rest-frame equivalent width (with respect to the continuum longward of the Lyman-$\alpha$ break) of EW$_0=12$\,\AA. Using a larger ``full-shutter" extraction aperture of height 5 pixels ($0.5\,\asec$) gives a significantly higher flux of $(2.30\pm 0.19)\times 10^{-18}$\,\flux, and the rest-frame equivalent width rises to EW$_0=18$\,\AA. The emission line flux of Lyman-$\alpha$ is about twice that of H-$\gamma$ (the strongest Balmer line we detect). From Case~B recombination and assuming no dust as found from the Balmer line ratio, Lyman-$\alpha$ would have about $50\times$ the line flux of H-$\gamma$, so it appears to be suppressed by about a factor of $26\times$ (i.e., $f_{\textrm{esc,Ly}\alpha}=0.038\pm 0.004$), presumably through resonant scattering effects. 

\begin{figure}
    \centering
    \includegraphics[width=\linewidth]{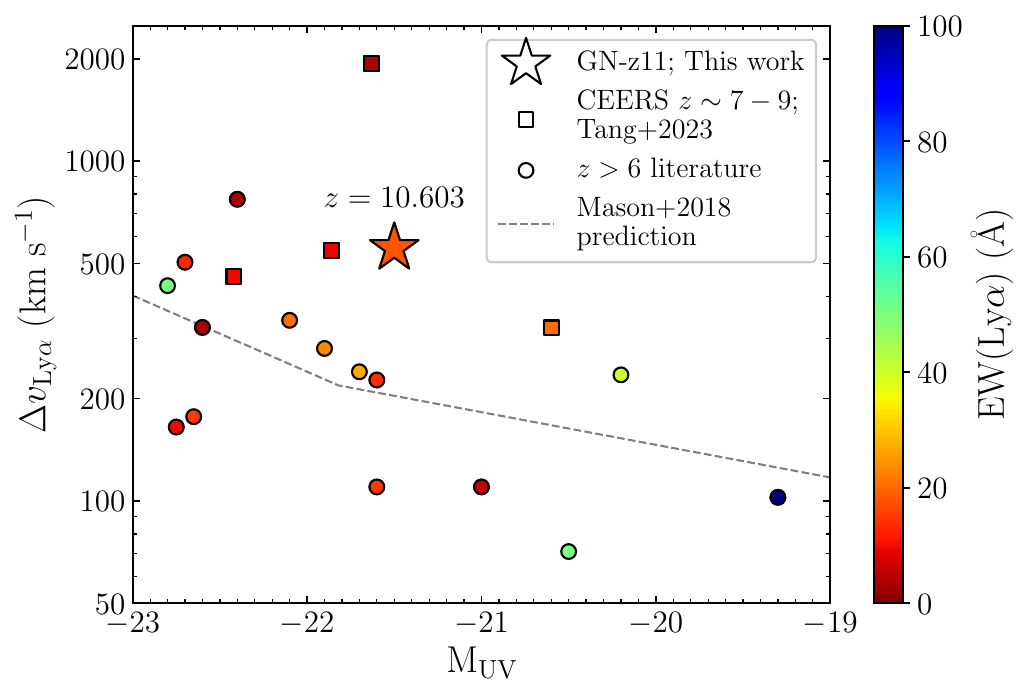}
\caption{Lyman-$\alpha$ velocity offset ($\Delta v_{\textrm{Ly}\alpha}$) versus M$_{\textrm{UV}}$ for GN-z11 (star) and other high-redshift galaxies, color coded by Lyman-$\alpha$ EW. We overplot data of $z>6$ galaxies with ground-based observations from the literature (\citealt{Cuby2003,Pentericci2011,Pentericci2016,Pentericci2018,Vanzella2011,Willott2013,Willott2015,Maiolino2015,Oesch2015,Stark2015,Stark2017,Furusawa2016,Knudsen2016,Carniani2017,Laporte2017,Mainali2017,Hashimoto2019}; see \citealt{Endsley2022b} and the Table 4 therein) in circles, and Ly$\alpha$ emitting galaxies at $z\sim7-9$ from CEERS NIRSpec observations \citep{Tang2023} in squares. Prediction of the correlation between Ly$\alpha$ velocity offset and M$_{\textrm{UV}}$ at $z=7$ from \citet{Mason2018a} is shown by the grey dashed line. GN-z11 has properties similar to the Ly$\alpha$ emitting galaxies at $z\sim7-9$.}
    \label{fig:Lya_offset}
\end{figure}

The discovery of Lyman-$\alpha$ emission at such high redshift is remarkable, given the expected highly neutral IGM at this epoch so much earlier than the end of reionization at $z\approx 6$ (\citealt{Fan2001}). However, perhaps this result is not so surprising when one considers the high rate of Lyman-$\alpha$ detection in luminous $7.5 < z < 9$ galaxies (\citealt{Zitrin2015,Oesch2015,Stark2017,Larson2023,Finkelstein2013,Roberts-Borsani2016,Jung2019,Song2106}) at redshifts prior to complete reionization. GN-z11 is a similarly luminous galaxy with $M_{\textrm {UV}}=-21.5$ so the effects that make Lyman-$\alpha$ detectable in such galaxies (see \citealt{Mason2018}) may be at play for GN-z11. 

There are two aspects of our Lyman-$\alpha$ observations that may explain the significant transmission of $f_{\textrm{esc,Ly}\alpha }=0.04$. 
Firstly, the large velocity offset of $555\pm 32$\,\kms and rest-frame equivalent width are similar to those measured in luminous $7.5 < z < 9$ galaxies with Lyman-$\alpha$ (Figure \ref{fig:Lya_offset}; see also \citealt{Tang2023}). Large velocity offsets are key to the escape of Lyman-$\alpha$ photons from galaxies in a highly-neutral IGM. Since the damping wing of the IGM and proximate HI will absorb photons close to the resonant frequency, photons that escape must resonantly scatter in the wings. Those that scatter far enough to the red may then be able to escape the system without being absorbed \citep{Dijkstra2014}. 
We note that the infall motion of the neutral gas around galaxies may also play a role (\citealt{Santos2004,Saudon2017,Weinberger2019,Park2021,Smith2022}). If the peculiar velocity of neutral gas near a Lyman-a source is infalling, then even photons redward of the Lyman-$\alpha$ will be resonantly absorbed, necessitating large velocity offsets of Lyman-$\alpha$ to facilitate its  escape.

Additionally, the intense star formation in luminous galaxies will be driving powerful and fast-moving outflows. Outflows on the far side of the galaxy may provide a redshifted medium from which the photons can backscatter to our line-of-sight with the required velocity offset. In this context, our observation of spatially-extended Lyman-$\alpha$ emission to the south-west suggests that if an outflow is present, it would extend in the north-east direction.

\subsection{Rest-frame UV properties of GN-z11}
\label{subsec:restuv}

From our low-dispersion prism spectra, where we have high S/N detection of the continuum, we measure a UV spectral slope of $\beta=-2.36\pm 0.10$ (over the range $\lambda_{\textrm{rest}}=1500-2600$\,\AA ), consistent with that of $\beta=-2.4$ reported from our NIRCam imaging in \citet{Tacchella2023}. We measure a luminosity of $M^{UV}_{AB}=-21.50\pm 0.02$ over the range $\lambda_{\textrm{rest}}=1400-1600$\,\AA\ (adopting a luminosity distance of 113,148.8\,Mpc from our chosen cosmology). This corresponds to a luminosity density of 
$L_{\nu}^{\mathrm{UV}}=1.7\times 10^{29}\,\textrm{erg}\,\textrm{s}^{-1}\,\textrm{cm}^{-2}\,\textrm{Hz}^{-1}$ 
around $\lambda_{\textrm{rest}}=1500$\,\AA . 

The rest-UV will be potentially affected by dust reddening. However, from the prism spectrum, we measure a Balmer line ratio using \hd/\hg\ of $0.53 \pm 0.06$, which is very close and within the errors of the intrinsic ratio of $0.55$ expected in an H\,\textsc{ii} region with electron density $n_e = 300$\,cm$^{-3}$ and temperature $T_e = 15,000$\,K, conditions expected in very high redshift galaxies \citep[e.g.][]{Curti2023, Katz2023_ero, Isobe2023}. We note that the wavelength baseline between \hd\ and \hg\ is short, so does not provide a large lever arm to quantify the dust attenuation accurately, and also that the \hd/\hg\ ratio of the fluxes from our 3-pixel medium grating spectrum is  $0.68\pm 0.20$, above the case B value of $0.55$ but consistent within the errors. Our full-SED \textsc{beagle} fits to the prism spectrum, assuming that this is a star-forming galaxy rather than an AGN, also suggest low attenuation ($A_V\approx 0.17$, only 70\% of which is from dust outside H\,{\sc ii} regions affecting nebular emission). Therefore, from the observed Balmer decrement and SED fitting we do not measure any considerable dust attenuation or reddening in the spectrum of GN-z11.

We now turn to the production and potential escape of ionizing photons in GN-z11. We can potentially constrain the escape fraction of ionizing photons ($f_{\textrm{esc}}$) by looking at how many produce local recombinations (as tracked by the hydrogen Balmer emission lines), compared with the total number produced.  \citet{Zackrisson2017} provide tracks of the rest-UV spectral slope ($\beta$, which is related to the hardness of the spectrum and hence the production of ionizing photons) and the equivalent width of \hb . Although we do not measure \hb\ in GN-z11, as it is beyond our spectral range, we can use the continuum slope we measure ($\beta\approx -2.4$) along with the flux in \hg\ and the case~B ratio (appropriate if $f_{\textrm{esc}}$ is low) of \hg/\hb=0.468 to estimate $\log(\mathrm{EW(H\beta)/\AA)}=1.8$. From the plots of \citet{Zackrisson2017}, this high \hb\ equivalent width suggests a low escape fraction of $f_{\textrm{esc}}\lesssim 0.1$. Our \textsc{beagle} SED fits also yield a  low $f_{\textrm{esc}}=0.03^{+0.05}_{-0.02}$. We do not see strong evidence for a ``reverse Balmer break'' in our NIRSpec spectra, which might be expected from nebular continuum if the escape fraction is indeed low, although in the best-fit \textsc{beagle} SED the reverse Balmer break is small. Conversely, we also see no evidence for a Balmer break that would be indicative of a moderately evolved stellar population. This is consistent with the very young age determined by the \textsc{beagle} SED fitting ($t\sim 19$\,Myr), with the light in the rest-frame UV and blue wavelengths we probe being completely dominated by a young stellar population.

Additional constraints on $f_{\textrm{esc}}$ could potentially be derived from the presence of \mgii\,$\lambda \lambda 2795,2802$ emission in the prism spectrum \citep[e.g.][]{Chisholm2020}. The considerable rest-frame EW of 12\,\AA\ for \mgii\ might suggest the presence of ionized channels in the galaxy, potentially facilitating the escape of ionizing radiation and Lyman-$\alpha$. Similar to Lyman-$\alpha$, the resonant nature of \mgii\ routinely causes strong absorption by low-ionization gas, while pure \mgii\ emission is thought to indicate a porous ISM \citep{Feltre2018, Henry2018, Witstok2021}. Following the predicted relationship in \citet{Witstok2021} between the strength of \mgii\ and the \neiii\ and \oii\ lines, we estimate an intrinsic \mgii\,$\lambda 2795$ flux of $3.1 \times 10^{-19}$\,\flux. This results in a \mgii\ escape fraction of $f_{\textrm{esc, \mgii}} \sim 60\%$ under the assumption of a typical doublet ratio of $F_{2795}/F_{2802} \approx 1.7$ (which itself depends on $f_{\textrm{esc, \mgii}}$ and the dust content; \citealt{Chisholm2020}). This is much larger than the estimated Lyman-$\alpha$ escape fraction, and also larger than the $f_\textrm{esc}$ inferred above from the equivalent width of the Balmer lines and the BEAGLE SED fitting. However, it has been suggested that \mgii\ escape could be more sensitive to dust rather than $f_\textrm{esc}$ particularly for galaxies in the optically thick regime, which may explain the discrepancy between the escape fractions measured from Balmer emission and \mgii\ \citep[e.g.][]{Katz2022}.

We now consider the production of ionizing photons, under the assumption of a low escape fraction $f_{esc}$ as discussed above, and compare these with the non-ionizing UV continuum detected. In case~B, $f(H\alpha)/f(H\gamma)=6.11$, and 45\% of recombinations result in an H$\alpha$ photon being emitted \citep{OsterbrockFerland2006}. We take the observed \hg\ line flux to be $1.2\times 10^{-18}$\,\flux\ from the 5-pixel extraction of the prism spectrum (since the agreement of the flux calibration with the NIRCam imaging is better than for the grating, Appendix~\ref{sec:compareNIRCam}). We make the assumption of no dust attenuation to obtain a hydrogen ionizing photon production rate of $N_\textrm{ion}=  8.8\times 10^{54}$\,photons\,s$^{-1}$. This gives an ionizing photon production efficiency of $\xi_\textrm{ion} = N_\textrm{ion} / L_{\nu}^\mathrm{UV} = 5.2\times 10^{25}$\,erg$^{-1}$Hz.
This value $\log\xi_\textrm{ion}=25.7$ agrees with that from the \textsc{beagle} SED fitting $\log\xi_\textrm{ion}=25.67\pm0.02$, and has  higher ionizing efficiency than galaxies at much lower redshift (e.g., \citealt{Chevallard2018} find $\log\xi_\textrm{ion}=25.2-25.8$ in extreme galaxies at $z\sim 0$, whereas \citealt{Bouwens2016} find $\log\xi_\textrm{ion}=25.3$ in sub-$L^*$ galaxies at $z=4-5$) but is comparable with that seen in $z\sim 7-8$ galaxies (e.g., \citealt{Tang2023} who find $\log\xi_\textrm{ion}=25.7-26.0$).

Although we have not ruled out an AGN component of GN-z11, we can place upper limits on the star formation rate based on the assumption that the observed line emission is powered solely by star formation.
From the rate of ionizing photons, we can estimate the star formation rate subject to assumptions about the star formation history and initial mass function (IMF) of stars. Using the \citet{Kennicutt1998} relation, $\mathrm{SFR}=1.08\times 10^{-53}(N_\textrm{ion}/\mathrm{s}^{-1})\,\textrm{M}_\odot \,\textrm{yr}^{-1}$ assuming a \citet{Salpeter1955} IMF, gives a star formation rate of $90\,\textrm{M}_\odot \,\textrm{yr}^{-1}$. For a \citet{Chabrier2003} IMF the star formation rate is $54\,\textrm{M}_\odot \,\textrm{yr}^{-1}$ from the \citet{Kennicutt1998} relation. Using the more recent H$\alpha$-based relation from \citet{Reddy2018}, which is more representative of the conditions found in galaxies at high redshifts, we obtain a star formation rate of $35\,\textrm{M}_\odot \,\textrm{yr}^{-1}$ again assuming \citet{Chabrier2003} IMF with an upper mass cut-off of $100\,\textrm{M}_{\odot}$.
Out \textsc{beagle} fit has a star formation rate of $\sim19\,\textrm{M}_\odot \,\textrm{yr}^{-1}$ for a Chabrier IMF with a higher upper-mass cut-off of $300\,\textrm{M}_{\odot}$. Using an upper-mass cut-off of $100\,\textrm{M}_{\odot}$ reduces the number of ionizing photons per unit SFR to 62\%, bringing the star formation rate to $31\,\textrm{M}_{\odot}\,\textrm{yr}^{-1}$, in agreement with the SFR derived using the \citet{Reddy2018} conversion.

We can also potentially use the rest-frame UV continuum to infer the star formation rate (or an upper limit on this, if there is an AGN contribution to the rest-frame UV). \citet{Kennicutt1998} give a relation $\mathrm{SFR}=1.4\times 10^{28} \times (L_{\nu}^\mathrm{UV}/ \mathrm{erg\,s^{-1}\,Hz^{-1})\,M_{\odot}\,yr^{-1}}$ for a Salpeter IMF, which would translate to a star formation rate of $24\,\textrm{M}_\odot \,\textrm{yr}^{-1}$ for GN-z11. However, this relation is probably inappropriate since it assumes constant star formation for 100\,Myr, and GN-z11 is likely much younger, so the UV luminosity will still be increasing even if star formation is constant, causing the star formation rate to be underestimated.

\begin{table}
    \centering
    \caption{Estimates of GN-z11 physical parameters derived from \textsc{beagle} SED fitting of the prism spectrum of Figure~\ref{fig:spectrum}}
    \begin{tabular}{l c }
    \hline
Parameter & GN-z11 \\
\hline
$\log(M / \textrm{M}_\odot)$  & $ 8.73^{+0.06}_{-0.06} $\\
$\psi / \textrm{M}_\odot \,\textrm{yr}^{-1}$ & $18.78^{+0.81}_{-0.69} $\\
$\log(t / \textrm{yr})$ & $7.27^{+0.19}_{-0.15} $\\
$\log(t_\textsc{m}/\textrm{yr})$ & $7.01^{+0.1}_{-0.07} $\\
$\log(Z_\textrm{neb} / Z_\odot)$ & $-0.92^{+0.06}_{-0.05}  $\\
$\log U_\textrm{S}$ & $-2.25^{+0.97}_{-0.87}  $\\
A$_V$ & $0.17^{+0.03}_{-0.03}$\\
$\log(\xi_\textrm{ion}/\textrm{erg}^{-1}\textrm{Hz})$ & $25.67^{+0.02}_{-0.02} $\\
$f_\textrm{esc}$ & $0.03^{+0.05}_{-0.02} $\\
    \hline
    \end{tabular}

    \label{tab:beagle}
     \textsc{beagle} SED fitting of the prism spectrum with the uncertainties giving the extent of the 1$\sigma$ credible regions: stellar mass ($M$, accounting for mass returned to the ISM through stellar winds and supernova explosions), star formation rate ($\psi$), maximum age of the stars ($t$), the mass-weighted age of stars ($t_\textsc{m}$), nebular metallicity ($Z_\textrm{neb}$), ionization parameter ($\log U_\textrm{S}$), $V$-band dust attenuation (A$_V$), ionizing photon production efficiency ($\xi_\textrm{ion}$) and escape fraction of H-ionizing photons ($f_\textrm{esc}$; see Appendix~\ref{sec:beaglefit} for details).

\end{table}

\subsection{ISM ionization and enrichment}
\label{subsec:ism}

In this section we use line ratio diagnostics to explore the ionization state and metal enrichment of the ISM, again under the assumption that the emission line fluxes are not dominated by an AGN contribution.
We detect a number of collisionally-excited metal lines, both of low ionization (\oii) and high ionization (including \niii, \neiii, \ciii), as well as Balmer lines from hydrogen recombination.
Our wavelength coverage does not extend to the widely-used \oiiiL5007, however we do have a robust detection of \neiiiL3869 which has a similar ionization potential. Hence, we consider the line flux ratio \neiiiL3869 / \oiiL3726, 3729 as a probe of ionization parameter ($U$) -- \neiii/\oii\ has been shown to track \oiii/\oii\ well \citep[e.g.][]{Levesque2014, Witstok2021}, which is the most widely-used indicator of $U$. We measure \neiii/\oii\ $=1.12\pm0.13$ from the 3-pixel grating extraction (\neiiiL3869 is blended in the prism), which is comparable to the redshift $z\sim5.5-9.5$ NIRSpec sample presented in \citet{Cameron2023} from our JADES survey, and also $z\gtrsim7$ galaxies observed in the CEERS survey \citep{Tang2023}.
Following the calibration set out in \citet{Witstok2021}, this corresponds to an ionization parameter of log $U=-2.03\pm0.04$. We find a similar value of $\log U=-2.25\pm 0.97$ from our \textsc{beagle} SED fitting.

We report a marginal detection of the \oiiiL 4363 line in our prism spectrum (partially blended with \hg; Figure~\ref{fig:emlines}), which has already been observed in a number of $z>7$ galaxies \citep[e.g.][]{Curti2023, Katz2023_ero}. Although this line can in theory be used to derive a $T_e$-based (`direct method') metallicity, the absence of \oiiiL 5007 from our data means we cannot measure the temperature with the standard approach. The \oiiiSFL 1660,1666 / \oiiiL 4363 ratio can also be used as a temperature diagnostic, but the low significance of the \oiiiL 4363 coupled with the marginal detection of \oiiiSFL 1660, 1666 in our grating spectrum means that any derived temperature would be highly uncertain. Thus, we instead consider using strong-line ratios to constrain the metallicity of GN-z11.
A widely-used metallicity indicator is R23 (the log of the ratio of \oii+\oiii\ to H$\beta$), but since \oiiiL5007 and H$\beta$ fall beyond our spectral coverage, we cannot measure this ratio. We instead consider an analogous ratio
of (\neiiiL3869 + \oiiL3727)/H$\delta$. All three of these emission lines are well detected in our grating spectra, and conveniently lie at very similar wavelengths which minimizes any uncertainties arising due to wavelength-dependent attenuation.
We measure a ratio of $\log_{10}(({\textrm \neiii} + {\textrm \oii})/H\delta) = 0.50\pm 0.07$ from the grating (3-pixel spectral extraction). Following the calibrations from \citet{Witstok2021} (which provides \oiii/\neiii\ $\approx15$ at the derived ionization parameter) and assuming $H\delta/H\beta=0.268$, this would be equivalent to $R23\approx0.85$.
These values place GN-z11 in fairly close alignment with the median values presented in the \citet{Cameron2023} sample;  their stacked spectra at $z\sim6$ ($z\sim8$) show $R23 = 0.88$ ($0.86$) and $\log$(\neiii/\oii)$=0.05$ ($0.04$). According to the binned average relationships presented in \citet{Nakajima2022}, this suggests a metallicity in the range $7.59<${\textrm 12+log(O/H)}$<7.76$, which corresponds to $0.08-0.12\,Z_{\odot}$ assuming a solar abundance of $12+\log(O/H)_\odot=8.69$.
Our \textsc{beagle} SED fitting yields a consistent value of $Z_{\textrm neb}=0.12\pm0.02 \,Z_\odot$.

\begin{figure}
    \centering
    \includegraphics[width=\linewidth]{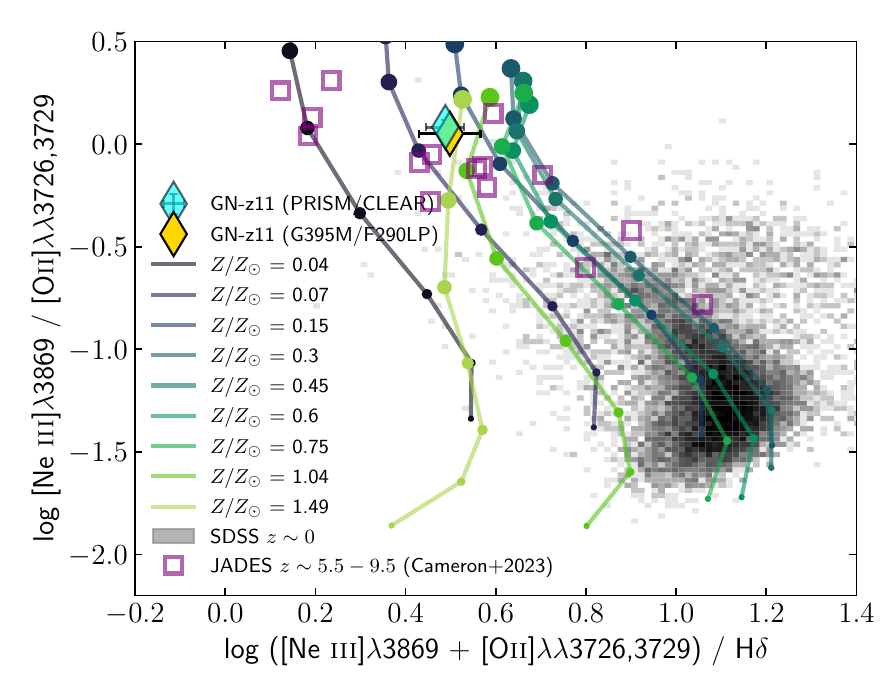}
    \caption{
    Line ratio diagram, showing (\neiii+\oii)/\hd\ vs. \neiii/\oii, featuring GN-z11 -- the yellow diamond denotes the line ratios derived from the medium-dispersion G395M grating, and the cyan diamond uses the low-dispersion prism spectrum (where we have corrected for the blending of \neiiiL3869 with  \hei\,$ \lambda$3889 using the flux ratio from the grating).
    The background grey 2D PDF shows the subset of SDSS galaxies with $0.03 < z < 0.1$ for which \neiiiL3869, \oiiL3726, 3729 and H$\delta$ are all detected with $S/N>5$.
    Purple squares show $z>5.5$ galaxies from \citet{Cameron2023} after adjusting the reported ratios to be in terms of H$\delta$ by assuming a fixed value of $H\delta/H\beta=0.268$.
    Solid lines show model grids from \citet{Gutkin2016}, plotted for nine different values of metallicity ($Z/Z_\odot =$ 0.04, 0.07, 0.15, 0.30, 0.45, 0.60, 0.75, 1.0, 1.5) indicated by the different colours, and seven values of ionization parameter, indicated by marker sizes, in steps of $0.5$ from $\log U=-4.0$ (smallest) to $\log U=-1.0$ (largest).
    }
    \label{fig:ratio_diagram}
\end{figure}

In Figure~\ref{fig:ratio_diagram} we compare our \neiii/\oii\ and (\neiii + \oii)/H$\delta$ measurements from GN-z11 (plotted separately from the prism and the grating data) with measurements from $z>5.5$ galaxies from \citet{Cameron2023}, $z\sim0$ galaxies from SDSS MPA-JHU catalogs\footnote{\url{https://www.sdss3.org/dr10/spectro/galaxy_mpajhu.php}} \citep{Aihara2011} and photoionization model grids from \citet{Gutkin2016}.
This line-ratio diagram is analogous to the widely used R23-O32 `ionization vs.\ excitation' diagram since, as described above, \neiii/\oii\ traces ionization and (\neiii + \oii)/H$\delta$ traces excitation of both the high- and low-ionization metal ions.

The \citet{Gutkin2016} models in Figure~\ref{fig:ratio_diagram} demonstrate the two-valued nature of (\neiiiL3869 + \oiiL3727)/H$\delta$ with metallicity.
Although the signal-to-noise ratio requirements significantly cut down the available SDSS sample, one can still see clear evidence of this two-valued relation.
The $z>5.5$ sample from \citet{Cameron2023} appears to follow an extrapolation of the low-metallicity (high-ionization) branch of this two-valued sequence.
We see that GN-z11 (diamond symbol) lies in good agreement with the sequence formed by these $z>5.5$ galaxies.
It falls between the \citet{Gutkin2016} $Z/Z_\odot =$ 0.07 and $Z/Z_\odot =$ 0.15 model lines, suggesting $12+{\textrm log(O/H)}\approx7.7$, and lies proximal to the model values with log $U=-2.0$, consistent with the empirical values derived above.

We now consider where GN-z11 might fall on the mass-metallicity relation (see \citealt{Maiolino2019} for a review).
The stellar mass estimated from \textsc{beagle} of $\log(M_*/M_{\odot})=8.73^{+0.06}_{-0.06}$ is consistent with that derived from our NIRCam photometry of $\log(M_*/M_{\odot})=9.1\pm^{0.3}_{0.4}$ presented in Tacchella et al.\ (2023), again assuming that the light is dominated by the stellar population rather than an AGN.
Our observed spectrum shows no evidence of a Balmer Break, and if the continuum is purely stellar, is dominated by a young stellar population.  It is possible a more stochastic star formation history would fit a higher stellar mass.
Comparing our metallicity and mass estimates for GN-z11 with the average reported for $8<z<10$ galaxies in \citet{Nakajima2023}
we find GN-z11 is offset to somewhat lower metallicity, albeit within the uncertainty quoted there. We note that the sample presented in that paper is still small and our understanding of the metallicities of galaxies at $z>8$ will no doubt continue to evolve significantly over the coming years.
We also note that the uncertainties on our derived metallicity are large.
In particular, we caution that the set of emission lines used to determine the metallicity presented here has not been robustly calibrated. The systematic uncertainties associated with this quoted metallicity are likely very high, so robust conclusions cannot be drawn from this about the evolution of the mass-metallicity relation.
Further work is needed to robustly calibrate shorter-wavelength metallicity diagnostics suitable for the study of $z>10$ galaxies with NIRSpec.

What is more puzzling is the strong \niii\ and \niv\ emission observed in the rest-frame UV, especially given the absence of a convincing detection of \oiiiSFL1660,1666 in our grating spectrum (although the blend with \heii\ is detected in the low-dispersion prism spectrum).
The \niiiL1748 emission line complex is not often seen in the spectra of star-forming galaxies, although it is detected in 2 of 44 galaxies in the low-redshift CLASSY survey \citep{Mingozzi2022}, including Mk996 \citep{James2009}. At intermediate redshifts, \niiiL1748  has been observed in stacks of rest-UV galaxy spectra at $z\sim 3$ \citep[e.g.][]{Saxena2022} and is weakly detected in SL2SJ021737-051329, a lensed  arc at $z = 1.84$ with low metallicity and high ionization \citep{Berg2018}.  However, \niiiL1748  is typically observed to be weaker than the nearby \oiiiL1660, 1666 lines.

This ratio of \niiiL1748 / \oiiiSFL1660, 1666 can be used to place constraints on nebular N/O abundance ratios \citep{Garnett1995}.
The conversion of \niii\ / \oiiiSF\ to N$^{++}$/O$^{++}$ depends on the electron temperature- and density-sensitive emissivities of these emission lines. However, the temperature and density dependence of these emission lines are remarkably similar. Adopting \niii\ / \oiiiSF\ $\gtrsim 2.6$ (see Table~\ref{tab:lines}), any adopted values of temperature $1 \leq T_e / 10^4 \text{K} \leq 55$ and density $2 \leq \log (n_e/\text{cm}^{-3}) \leq 10$ results in emissivities that imply $0.72 \leq \text{N}^{++}/\text{O}^{++} \leq 1.09$, significantly higher than the solar N/O value of 0.14\footnote{Emissivity calculations were performed with {\sc pyneb} \citep{Luridiana2015}.}

The measured $\text{N}^{++}/\text{O}^{++}$ may over-estimate the total N/O if there is a significant fraction of oxygen in other ionisation states. The second and third ionisation energies of nitrogen (29.6 and 47.4 eV) are milder than those of oxygen (35.1 and 54.9 eV). Assuming a simply multi-zone model of the ionisation structure of the ISM, this implies that $\text{N}^{++}/(\text{O}^{+} + \text{O}^{++})$ should be a lower limit on the total N/O. However, from detections of \oiiL 3727 and \oiiiL 4363 emission for GN-z11, if we assume the \oii\ emission arises from gas with density below the critical density of \oiiL 3727 (i.e. $n_e \lesssim 10^4 K$), we get an approximate lower limit of $\text{O}^{++}/\text{O}^{+} \gtrsim 1.5$, consistent with the expectation of a highly ionised ISM. Thus, even if we conservatively assume $\text{O}^{++}/\text{O}^{+} = 1.5$, this only lowers the inferred N/O by a factor of 0.6, implying a lower limit on the total nitrogen to oxygen abundance ratio of N/O $> 0.43$, or $log_{10} (\text{N}/\text{O}) > -0.36$, more than two times higher than the solar abundance ratio.
This would appear quite unusual with respect to $z\sim0-2$ galaxies \citep[e.g.][]{PerezMontero2009, HaydenPawson2022}, and strongly inconsistent with canonical chemical evolution models (see \citealt{Maiolino2019} or \citealt{Kobayashi2022} for reviews).

We note that this simple calculation is independent of the excitation source (i.e. stellar photoionisation or AGN). However, we cannot rule out the scenario in which only a small fraction of the gas in GN-z11 is highly nitrogen enriched, but that this gas is extremely luminous in emission and dominates the global spectrum.
More detailed modelling of the ionisation states of nitrogen and oxygen throughout the ISM of GN-z11 would be required to derive a more precise value of N/O, which is beyond the scope of this paper.

We also detect the [\ciii$\lambda1907+$\,\ciii$\lambda 1909$ line in our G235M spectrum. The \ciii\ line has been much more widely observed in star-forming galaxies at high redshift \citep[e.g.][]{Saxena2022, ArellanoCordova2022, Jones2023}, and its presence does not necessarily point to unusual C/O abundance ratios \citep{ArellanoCordova2022, Jones2023}.

In summary, the emission line ratios measured for GN-z11 suggest a very high ionization parameter and low oxygen abundance in the vicinity of 10~\% solar, broadly in line with findings from galaxies at $z\sim6-10$ \citep{Cameron2023, Sanders2023, Mascia2023, Nakajima2023, Tang2023}. However, the detection of strong \niii\ emission suggests unexpected abundance patterns, which may have deeper implications for chemical enrichment histories.

\section{Conclusions}
\label{sec:conclusions}

We present JWST/NIRSpec spectroscopy of one of the most luminous galaxies at $z>10$. GN-z11 is in the GOODS-North field and had previously been identified as a Lyman break galaxy candidate by \citet{Oesch2015}, with a tentative redshift of $z=11.1$ from a continuum break in slitless HST/WFC3 spectroscopy \citep{Oesch2016}. We see numerous emission lines and a strong Lyman-$\alpha$ break in our NIRSpec spectroscopy, and we unambiguously measure the redshift to be $z=10.603$. 
Our grating spectrum reveals Lyman-$\alpha$ in emission, making it the first object at $z>9$ with confirmed Lyman-$\alpha$ emission. The rest-frame equivalent width is $W_0=18$\,\AA. The emission is offset 555\,\kms redward of the systemic redshift and spatially extended. These properties are consistent with models of Lyman-$\alpha$ backscattering off the far side of galactic scale outflows. 

The NIRSpec spectrum of GN-z11 is remarkably rich with emission lines, enabling us to study the ISM properties at $z>10$. Based on the high \neiii/\oii\ ratio we infer a high ionization parameter ($\log(U)>-2.0$). We report a significant detection of the very rarely-seen \niiiL$1748$ line, which could suggest unusually high N/O ratios. While some high ionization lines are detected, the \heii\,$\lambda1640$ and \civ$\,\lambda1550$ lines, which are typically associated with photoionization due to AGN, are weak. Although we cannot conclusively rule our the contribution of an AGN, if this galaxy is indeed powered by star formation then the Balmer emission lines and UV continuum suggest a current star formation rate of $\sim 30\,M_{\odot}$\,yr$^{-1}$ and low dust attenuation.

We have presented a very high signal-to-noise spectrum of a galaxy at $z>10$, showing continuum and line emission, highlighting the  power of our JADES observations to not only measure redshifts but to do detailed studies of the physical and chemical properties of galaxies formed within the first few hundred million years of the Big Bang. 
  
\begin{acknowledgements}

AJB, AS, AJC, GCJ, JC, and IEBW acknowledge funding from the "FirstGalaxies" Advanced Grant from the European Research Council (ERC) under the European Union’s Horizon 2020 research and innovation programme (Grant agreement No. 789056).
ECL acknowledges support of an STFC Webb Fellowship (ST/W001438/1).
The Cosmic Dawn Center (DAWN) is funded by the Danish National Research Foundation under grant no.140.
RS acknowledges support from a STFC Ernest Rutherford Fellowship (ST/S004831/1).
RM, JW, MC, FDE, JS, TJL, LS, and WMB acknowledge support by the Science and Technology Facilities Council (STFC) and by the ERC through Advanced Grant 695671 "QUENCH". RM also acknowledges funding from a research professorship from the Royal Society.
JW also acknowledges funding from the Fondation MERAC.
This research is supported in part by the Australian Research Council Centre of Excellence for All Sky Astrophysics in 3 Dimensions (ASTRO 3D), through project number CE170100013.
FS, EE, DJE, BDJ, MR, BER, IS, and CNAW acknowledge a JWST/NIRCam contract to the University of Arizona NAS5-02015.
DJE is also supported as a Simons Investigator.
SC acknowledges support by European Union’s HE ERC Starting Grant No. 101040227 - WINGS.
SA, BRDP, and MP acknowledges support from the research project PID2021-127718NB-I00 of the Spanish Ministry of Science and Innovation/State Agency of Research (MICIN/AEI).
H\"U gratefully acknowledges support by the Isaac Newton Trust and by the Kavli Foundation through a Newton-Kavli Junior Fellowship.
Funding for this research was provided by the Johns Hopkins University, Institute for Data Intensive Engineering and Science (IDIES).
RB acknowledges support from an STFC Ernest Rutherford Fellowship [grant number ST/T003596/1].
MP also acknowledges support from the Programa Atracci\'on de Talento de la Comunidad de Madrid via grant 2018-T2/TIC-11715.
LW acknowledges support from the National Science Foundation Graduate Research Fellowship under Grant No. DGE-2137419.
DP acknowledges support by the Huo Family Foundation through a P.C. Ho PhD Studentship.

This work is based [in part] on observations made with the NASA/ESA/CSA James Webb Space Telescope. The data were obtained from the Mikulski Archive for Space Telescopes at the Space Telescope Science Institute, which is operated by the Association of Universities for Research in Astronomy, Inc., under NASA contract NAS 5-03127 for JWST. These observations are associated with program \#1181.
      
\end{acknowledgements}

%
%

\bibliographystyle{aa}
\bibliography{biblio}

\newpage

\begin{appendix}

\section{\textsc{beagle} fit to the prism spectrum}
\label{sec:beaglefit}

\begin{figure*}
    \centering
\includegraphics[width=\linewidth]{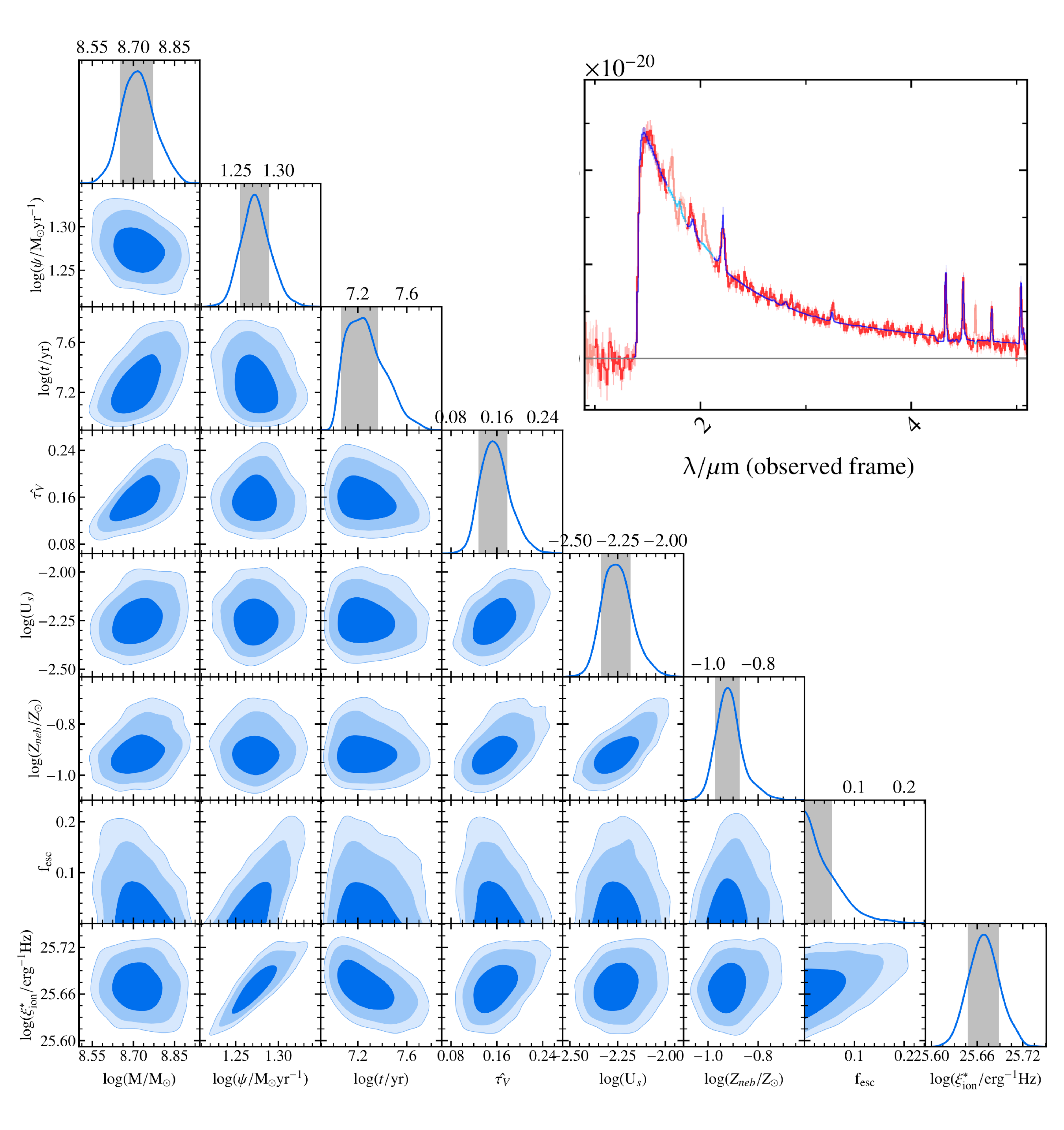}
    \caption{Triangle plot showing the 2D (1D on diagonal) posterior probability distributions for the derived stellar mass ($M$, accounting for mass returned to the ISM through stellar winds and supernova explosions), star formation rate ($\psi$), maximum age of the stars ($t$), effective $V$-band attenuation optical depth ($\hat{\tau_v}$ = A$_\textrm{V}/1.086$), ionization parameter ($\log U_\textrm{S}$), nebular metallicity ($Z_\textrm{neb}$), escape fraction of H-ionizing photons ($f_\textrm{esc}$) and ionizing photon production efficiency ($\xi_\textrm{ion}$).  The contours in the 2D posterior plots show the 1, 2 and 3$\sigma$ credible regions in light, medium and dark blue, respectively. The inset shows the resulting fit to the prism spectrum, with the spectrum and $1\sigma$ standard errors shown as red line and shaded region respectively, and 1$\sigma$ range of fitted model spectra in blue.  Regions that are masked in the spectrum are shown as fainter red (data) and fainter blue (model) shaded regions.}
    \label{fig:BEAGLEmarginal}brew install ilmbase
\end{figure*}

The prism spectrum was fitted to with \textsc{beagle} \citep{Chevallard2016} using the updated \cite{bc03} stellar models as described in \cite{VidalGarcia2017}, with a \cite{Chabrier2003} initial mass function with upper and lower mass cutoffs of 300M$_\odot$ and 0.1M$_\odot$, respectively.  The results are given in Tab. \ref{tab:beagle}. Nebular line+continuum emission is modelled following \cite{Gutkin2016}.  Treatment of the instrument line spread function is described in \cite{Curtis-Lake2022}.  We fit varying all stellar and nebular parameters (metallicity, ionization parameter, dust-to-gas mass ratio), employ a delayed exponential star formation history with recent 10 Myr of constant star formation that can vary independently.  Finally, we model attenuation by dust using the \cite{CF00} two-component dust law.  We also allow for a damping wing for the neutral intergalactic medium following the prescription described in \cite{Curtis-Lake2022}, though we do not report the values here as they will be affected by the Ly$\alpha$ flux which is unresolved in the PRISM spectrum. The resulting spectral fit and derived parameters are shown in Fig. \ref{fig:BEAGLEmarginal} and reported in Table~\ref{tab:beagle}.  The \cite{Gutkin2016} nebular models employ a relation between N/O and O/H abundances, which gives low N/O abundance at low metallicity.  As such, our fit does not reproduce the rest-frame UV Nitrogen lines.  We therefore mask them to prevent their presence affecting the fit to the rest-frame UV continuum.  We also mask the region around \civ, which shows an offset in the emission from the expected wavelength.

\section{\texorpdfstring{$R\sim1000$}{R1000} grating 1D and 2D spectra}

\begin{figure*}[h]
    \centering
    \includegraphics[width=0.9\linewidth]{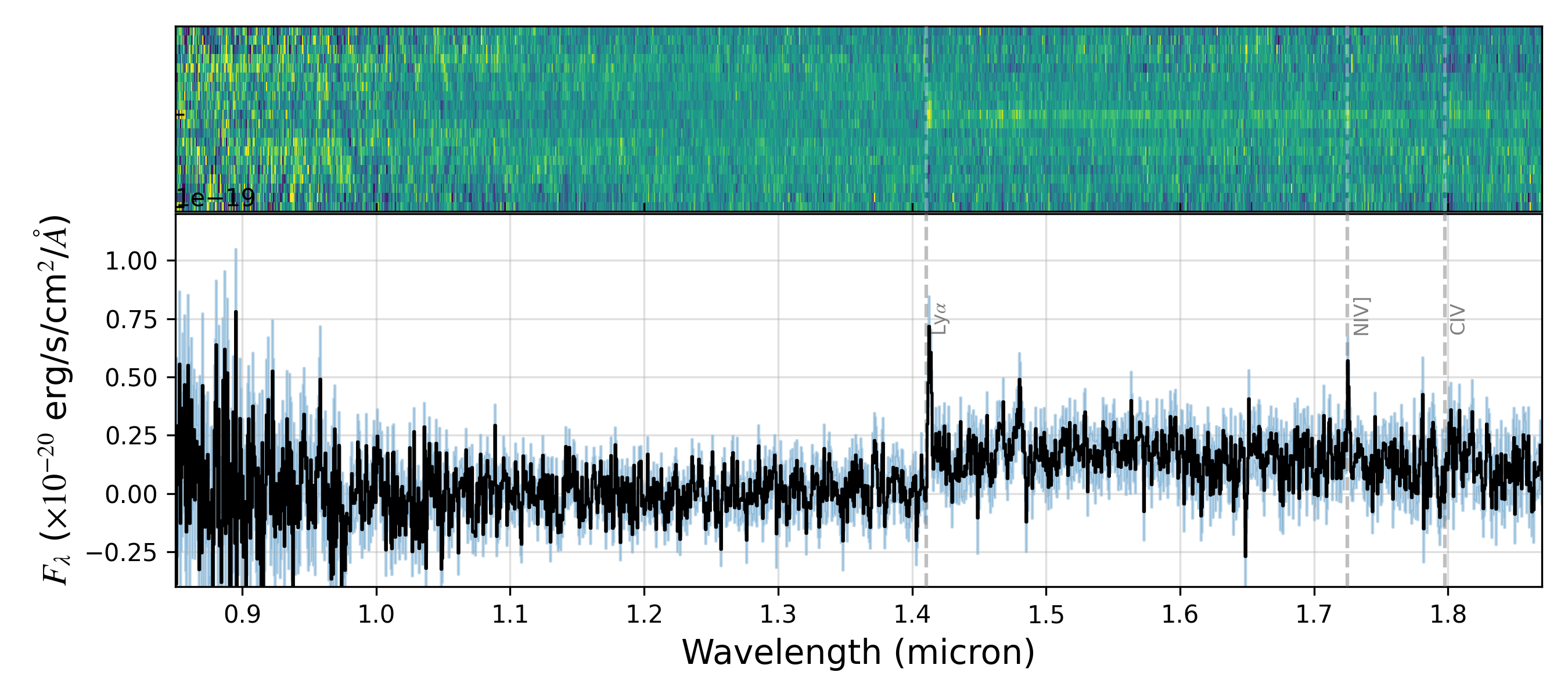}

    \includegraphics[width=0.9\linewidth]{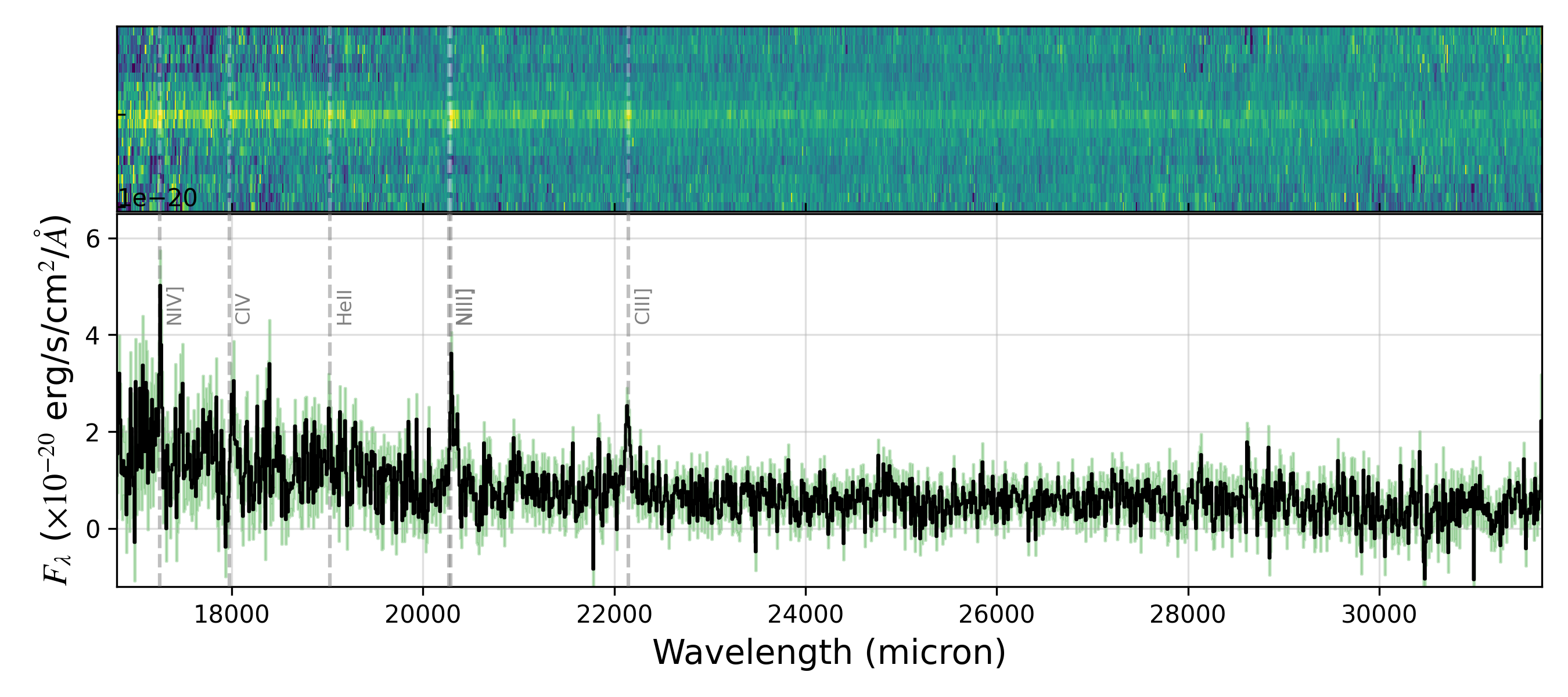}

    \includegraphics[width=0.9\linewidth]{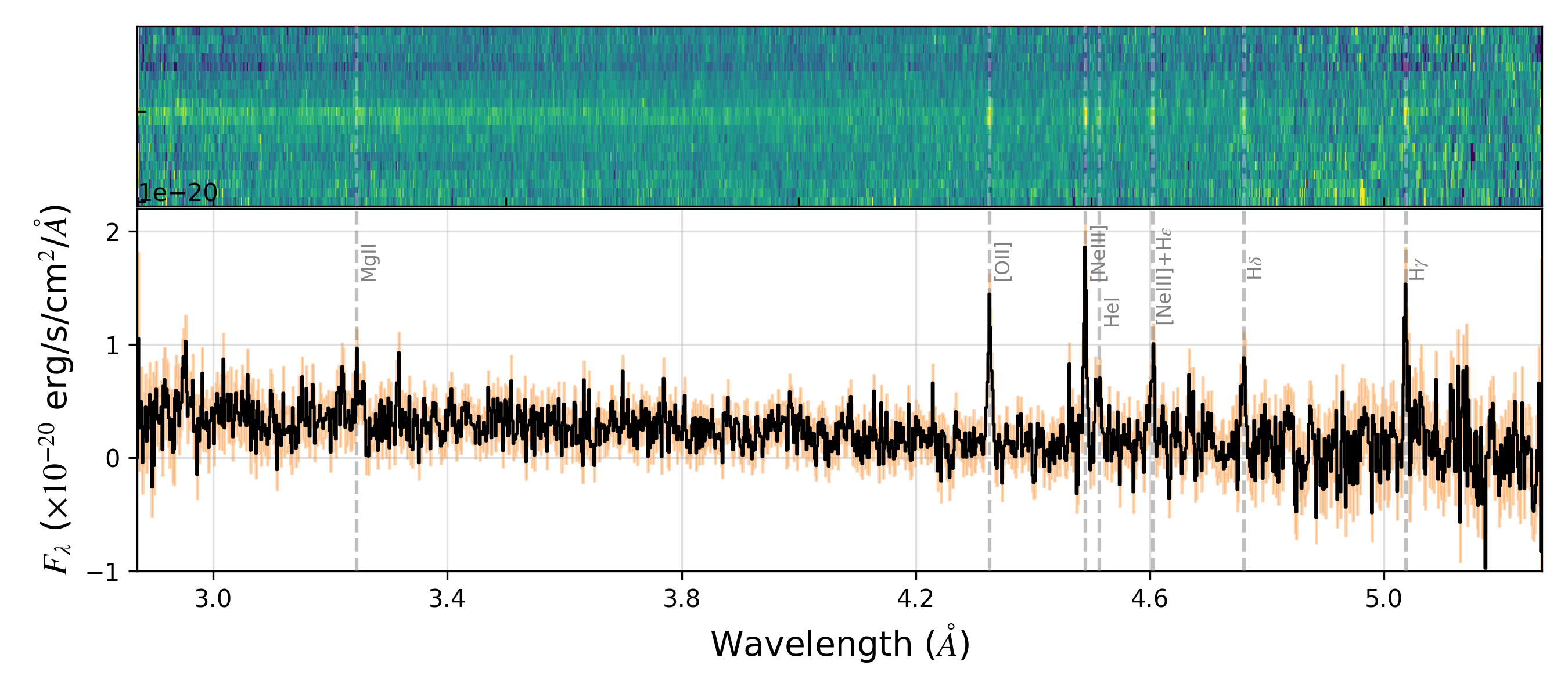}
    \caption{Full coverage of the 2D and 1D spectra from the medium resolution G140M (top), G235M (middle) and G395M (bottom) gratings. The main emission lines observed in the spectra have been marked.}
    \label{fig:gratings}
\end{figure*}

Figure~\ref{fig:gratings} shows the full NIRSpec medium-resolution grating spectra with detected emission lines marked.

\section{Comparison of NIRSpec flux calibration to NIRCam imaging}
\label{sec:compareNIRCam}

As a check on our spectroscopic flux calibration, we compare the GNz-11 fluxes derived from the NIRSpec spectroscopy with the NIRCam fluxes across several filters reported in \citet{Tacchella2023}. Given the small spatial size of the source, and the narrow ($0\farcs2$-wide) NIRSpec microshutters, we use the $0\farcs2$-diameter NIRCam aperture photometry (corrected to total magnitudes assuming a point source) reported in \citet{Tacchella2023} and our 3-pixel ($0\farcs 3$-wide) NIRSpec spectral extraction (again corrected for slit losses to approximate total flux assuming a point source at the location of GN-z11 with the microshutter).  Each extracted spectrum was converted to photons per unit wavelength (since NIRCam detects photons, not energy), and the flux density integrated over the bandpass of each filter, weighting by the filter transmission curve. We then computed the brightness (in nJy) of a source with a spectrum uniform in $f_{\nu}$ (i.e. flat in AB magnitudes) which would produce the same integrated flux, so as to compare with the quoted fluxes for NIRCam on the AB system (see Table~\ref{tab:compareNIRCam}).  

\begin{table}
    \centering
    \caption{Comparison of the NIRCam photometry of GN-z11 with the NIRSpec spectroscopy.}
    \begin{tabular}{l c c c l}
    \hline
Filter & Prism flux & Grating flux & Grating \\
& nJy & nJy & \\
\hline
F090W      $-1.0 \pm 1.7$       &   $ 1.4  \pm   0.9     $    &  $  -3.0   \pm    3.8$  & G140M \\
F115W      $1.2   \pm   1.5$    &   $  0.6  \pm   0.9    $    &  $   -3.6   \pm    1.7$	& G140M \\
F150W      $99.2   \pm   1.6$   &   $ 100.0  \pm    1.1  $    &  $103.0   \pm    3.3$	& G140M \\
F200W      $135.2  \pm   1.5$   &   $133.8  \pm    1.0   $    & $139.7   \pm    2.9$	& G235M \\
F277W      $112.0  \pm   1.0$   &   $115.1  \pm   0.9    $    & $132.8   \pm    3.5$	& G235M \\
F335M      $107.2  \pm   1.9$   &   $113.0  \pm    1.3   $    &  $127.2  \pm     4.1$	& G395M \\
F356W      $106.7  \pm   1.0$   &   $112.5  \pm    1.0   $    &  $124.3  \pm     2.9$	& G395M \\
F410M      $109.9  \pm   1.4$   &   $112.8  \pm    1.8   $    &  $108.3  \pm     5.0$	& G395M \\
F444W      $121.0  \pm   1.3$   &   $118.68  \pm    1.5  $    &  $ 114.7   \pm    4.5$	& G395M \\
    \hline
    \end{tabular}

    \label{tab:compareNIRCam}
    
    The NIRCam photometry of GN-z11 in a $0\farcs2$-diameter aperture (column 2) from \citet{Tacchella2023} is compared with that inferred from the spectroscopy in the low-dispersion prism (column 3) and medium-dispersion gratings (flux in column 4, and grating which overlaps that filter in column 5). The spectroscopic measurements use the 3-pixel extraction.
    
\end{table}

\begin{figure*}[h]
    \centering
    \includegraphics[width=0.9\linewidth]{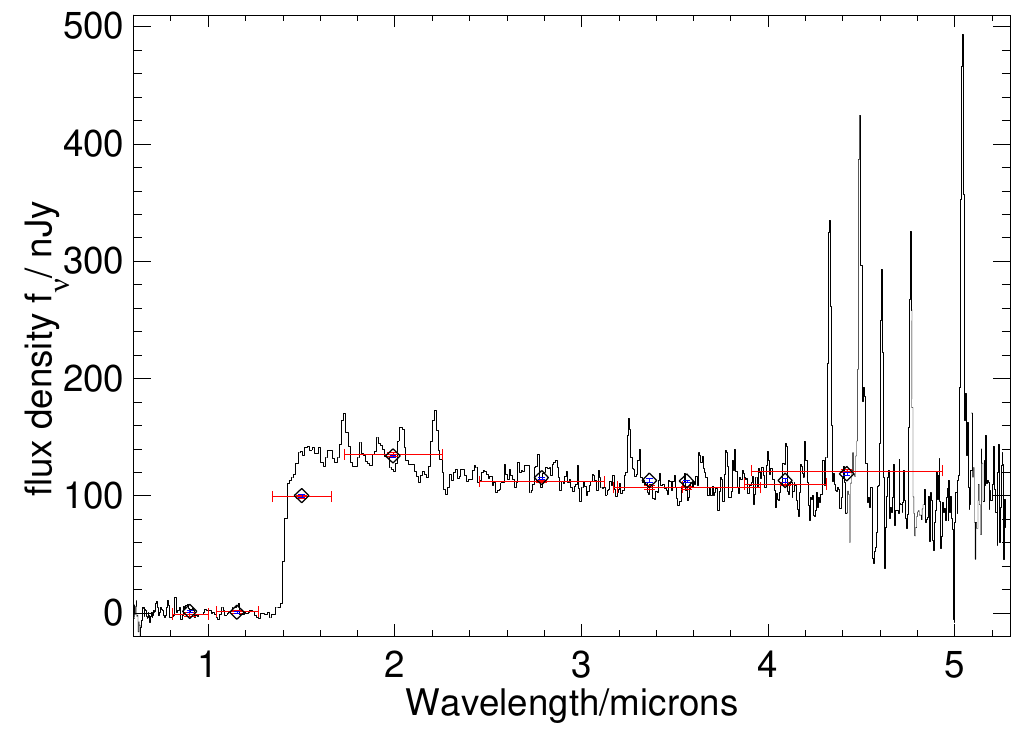}
    \caption{The NIRSpec low-dispersion prism spectrum of GN-z11 (3-pixel extraction), compared with the NIRCam photometry in different filters (red error bars, denoting the flux uncertainty and the wavelength span of the filter bandpass). The large diamond symbols denote the flux from the NIRSpec spectrum integrated over the filter response curve, with the small blue error bars within these diamond symbols.}
    \label{fig:compareNIRCamPrism}
\end{figure*}

As can be seen from Figure~\ref{fig:compareNIRCamPrism}, the agreement in flux between the low-dispersion prism spectrum and the NIRCam photometry is excellent, with most filters agreeing with the spectral fluxes within the nominal error bars (the agreement is generally within $5$\%).
The flux calibration of the grating spectra is less accurate.
The grating spectra show significantly higher fluxes than NIRCam (or the prism spectrum) at $2.5<\lambda <4\,\mu$m at the $\approx 15-20$\% level, greater than the nominal uncertainties in the photon statistics, although at other wavelengths the agreement is better. The exact origin of this is unclear, but the flux calibration of the gratings may be less good, or there may be background subtraction issues affecting the measured flux.

\begin{figure*}[h]
    \centering
    \includegraphics[width=0.9\linewidth]{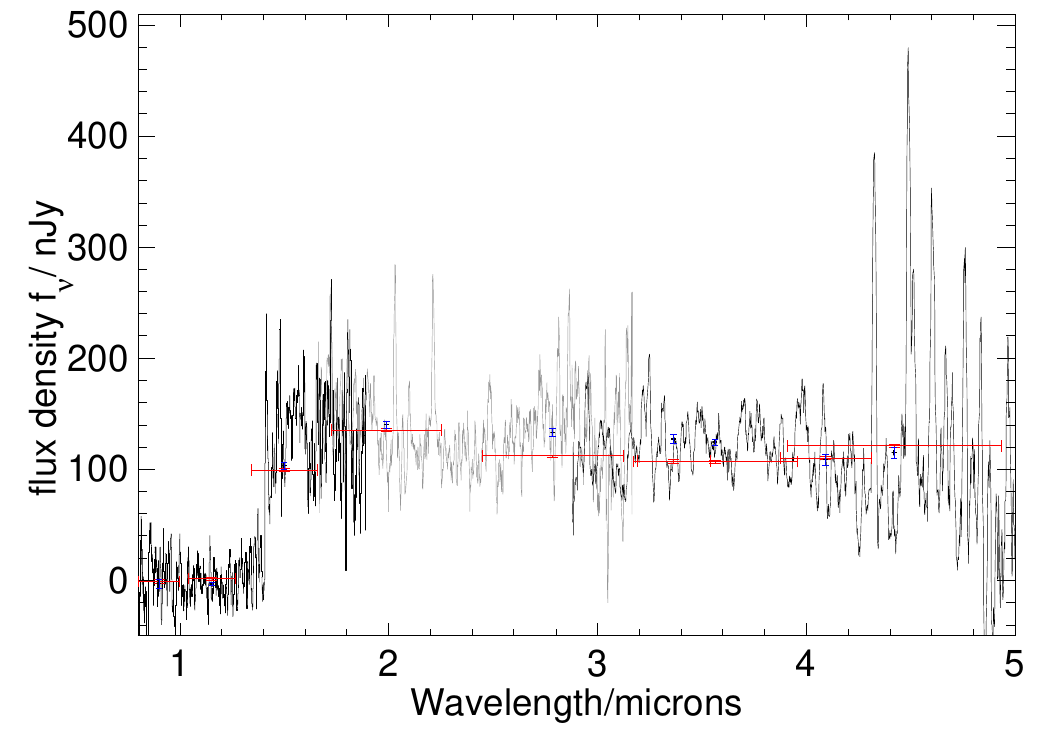}
    \caption{The NIRSpec medium-dispersion grating spectrum of GN-z11 (3-pixel extraction, smoothed in the spectral direction with a 11-pixel boxcar), compared with the NIRCam photometry in different filters (red error bars, denoting the flux uncertainty and the wavelength span of the filter bandpass). The light grey spectrum is the G235M, with the G140M and G395M grating spectra in black. The `$+$' symbols and blue error bars denote the flux from the NIRSpec spectrum integrated over the filer response curve.}
    \label{fig:compareNIRCamGratings}
\end{figure*}

\end{appendix}

\end{document}